\pdfoutput=1 
\documentclass[12pt]{iopart}

\usepackage{iopams}  
\expandafter\let\csname equation*\endcsname\relax

\expandafter\let\csname endequation*\endcsname\relax

\usepackage{graphicx}
\usepackage{epstopdf}
\usepackage{amsmath}
\usepackage{braket}
\usepackage{caption}
\usepackage{subcaption}
\usepackage[colorlinks=false,hidelinks]{hyperref}
\usepackage[capitalise,nameinlink,sort&compress]{cleveref}
\usepackage{cite}
\usepackage{algorithm}
\usepackage{algpseudocode}

\usepackage[colorinlistoftodos]{todonotes}

\renewcommand{\etal}{\textit{et al.}}

\usepackage{mdframed}
\newcommand{\rqbox}[1]{
    \begin{mdframed}
    [
        innerbottommargin=.3\baselineskip,
        rightmargin=0em,
        leftmargin=0em
    ]
    #1
    \end{mdframed}
}

\newcommand{\researchquestion}[1]{
    \rqbox{\textbf{Main RQ:} \textit{#1}}
}

\newcounter{rq_counter}
\newcommand{\subresearchquestion}[1]{
    \addtocounter{rq_counter}{1}
    \rqbox{\textbf{SRQ\,\arabic{rq_counter}:} \textit{#1}}
}

\newcommand{\ketbra}[2]{\vert #1 \rangle\langle #2 \vert}

\newcommand{\superop}{\mathcal{S}}

\begin{document}

\title[Circuit cutting in QAOA for the MaxCut problem on NISQ devices]{Investigating the effect of circuit cutting in QAOA for the MaxCut problem on NISQ devices}

\author{Marvin Bechtold, Johanna Barzen, Frank Leymann, Alexander Mandl, Julian Obst, Felix Truger and  Benjamin Weder}

\address{Institute of Architecture of Application Systems, University of Stuttgart, Universitätsstraße 38,
70569 Stuttgart, Germany}
\ead{firstname.lastname@iaas.uni-stuttgart.de}
\vspace{10pt}
\begin{indented}
\item[]July 2023
\end{indented}

\begin{abstract}
\frenchspacing
Noisy Intermediate-Scale Quantum~(NISQ) devices are restricted by their limited number of qubits and their short decoherence times.
An approach addressing these problems is quantum circuit cutting.
It decomposes the execution of a large quantum circuit into the execution of multiple smaller quantum circuits with additional classical postprocessing.
Since these smaller quantum circuits require fewer qubits and gates, they are more suitable for NISQ devices.
To investigate the effect of quantum circuit cutting in a quantum algorithm targeting NISQ devices, we design two experiments using the Quantum Approximate Optimization Algorithm~(QAOA) for the Maximum Cut~(MaxCut) problem and conduct them on state-of-the-art superconducting devices.
Our first experiment studies the influence of circuit cutting on the objective function of QAOA, and the second evaluates the quality of results obtained by the whole algorithm with circuit cutting.
The results show that circuit cutting can reduce the effects of noise in QAOA, and therefore, the algorithm yields better solutions on NISQ devices.
\end{abstract}

%
\noindent{\it Keywords\/}: Circuit cutting, NISQ, QAOA, MaxCut, quantum computing

%
\submitto{Quantum Science \& Technology}
%
\maketitle
%
%

\section{Introduction}
\frenchspacing
Quantum computing promises to solve problems that are intractable for classical high-performance computers~\cite{Cao2018,Cao2019}.
However, the current \textit{Noisy Intermediate-Scale Quantum~(NISQ)} devices have a limited number of qubits, and computations on them are flawed due to decoherence of the quantum state, inaccuracy of implemented gates, and erroneous measurements~\cite{Preskill2018,Salm2020}.
These errors accumulate during the computation, and thus, the error rate of a NISQ device restricts the width and depth of executable circuits~\cite{Leymann2020a,Salm2020}.
Due to these deficiencies, NISQ devices demand hybrid algorithms that combine shallow quantum circuits and classical computing to leverage the advantages of both~\cite{Leymann2020a}.
This includes the class of \textit{Variational Quantum Algorithms (VQAs)} that repeatedly run a parameterized quantum circuit, the so-called \textit{ansatz}, on a quantum device and utilize a classical optimizer for parameter optimization~\cite{Cerezo2021}.
These algorithms can tolerate certain amounts of noise during their computations and are considered a promising approach to achieve quantum advantage on NISQ devices. 
A prominent VQA for solving combinatorial optimization problems is the \textit{Quantum Approximate Optimization Algorithm (QAOA)}~\cite{Farhi2014}.
Even for circuits of small depth, QAOA cannot generally be simulated efficiently on any classical device~\cite{Farhi2016}.

Still, the number of available qubits is too small~\cite{Dalzell2018}, and the noise of the current NISQ devices is too high~\cite{Wang2021,Harrigan2021,Alam2019} to exploit the potential of QAOA to achieve quantum advantage.
Consequently, strategies that reduce the size of quantum circuits, i.e., their width and depth, may help to overcome these challenges.
One method to address this is quantum circuit cutting~\cite{Bravyi2016,Peng2019,Mitarai2021}.
The main idea is that a large quantum circuit that requires many qubits can be cut into smaller subcircuits requiring fewer qubits.
In a postprocessing step, a classical computer produces the result of the original quantum circuit by combining the results obtained from running the subcircuits.
This allows the evaluation of large quantum circuits using small quantum computers and additional classical postprocessing.
Moreover, cutting circuits into subcircuits can also reduce their depth~\cite{Peng2019}.
Since shallower circuits are less susceptible to noise, they are better suited for NISQ devices~\cite{Leymann2020a}.
However, the potential of circuit cutting in QAOA is an open question.
\looseness=-1

Therefore, we approach this question by investigating to what extent circuit cutting can improve the results of QAOA in the presence of noise.
To answer this, we design two experiments that apply circuit cutting in QAOA for the \textit{Maximum Cut~(MaxCut)} problem and execute them on state-of-the-art NISQ devices.
The first experiment evaluates how the computed objective function on NISQ devices changes when circuit cutting is applied in QAOA.
The second experiment studies how these changes in the objective function influence the approximated solution of QAOA.
Therefore, we classically optimize the parameters of the QAOA circuits with and without circuit cutting and compare the results achieved on NISQ devices.

The remainder of the paper is structured as follows. 
Next, \Cref{sec:background} introduces the relevant background, and \Cref{sec:relatedwork} discusses related work.
Based on this, \Cref{sec:motivation} motivates the problem and refines the research question.
Following this, \Cref{sec:design} describes the research design, and \Cref{sec:results} presents the results.
Afterward, \Cref{sec:discussion} discusses the findings, and \Cref{sec:conclusion} concludes the work.

\section{Background}\label{sec:background}
This section briefly highlights the fundamentals of QAOA and its application to the MaxCut problem. 
Afterward, we introduce quantum circuit cutting and present the later used technique.

\subsection{QAOA and the MaxCut problem}
QAOA enables solving combinatorial optimization problems such as the MaxCut~\cite{Farhi2014} and the maximum independent set problem~\cite{Farhi2020}.
The goal is to find a bitstring $\boldsymbol{z}=(z_{1},...,z_{n})\in\{0,1\}^{n}$ that maximizes an objective function $\mathcal{C}(\boldsymbol{z})$ which is encoded on the diagonal of the cost Hamiltonian $H_{C}$ such that $H_{C}\ket{\boldsymbol{z}} = \mathcal{C}(\boldsymbol{z})\ket{\boldsymbol{z}}$.
Thus, the optimization problem can be solved by finding the eigenvector $\ket{\boldsymbol{z}}$ of $H_{C}$ with maximal eigenvalue 
\begin{equation}
    \max_{\ket{\boldsymbol{z}}} \bra{\boldsymbol{z}} H_{C} \ket{\boldsymbol{z}} = \mathcal{C}_{\max}.
\end{equation}

To this end, QAOA employs an ansatz that applies alternately $p$ times the cost operator $U(H_{C}, \gamma_{l}) = \exp{(-i\gamma_{l}H_{C})}$ and mixing operator $U(H_{B}, \beta_{l}) = \exp{(-i\beta_{l}H_{B})}$ to the initial state $\ket{+}^{\otimes n}$~\cite{Farhi2014}.
Herein, $H_{B} = \sum_{i}X_{i}$, where $X_{i}$ is the Pauli X matrix applied to the $i$-th qubit.
Thus, the result state of the QAOA ansatz is:

\begin{equation}
    \ket{\psi(\boldsymbol{\beta}, \boldsymbol{\gamma})} = \left(\prod_{l = 1}^{p}U(H_{B}, \beta_{l})U(H_{C}, \gamma_{l}) \right)\ket{+}^{\otimes n}
\end{equation}

\noindent
where $\boldsymbol{\beta}=(\beta_{1}, ..., \beta_{p})$ and $\boldsymbol{\gamma}=(\gamma_{1}, ..., \gamma_{p})$ are variational parameters.
A classical optimizer updates the parameters $\boldsymbol{\beta}, \boldsymbol{\gamma}$ to maximize the expectation value of the observable $H_{C}$ on the ansatz $\ket{\psi(\boldsymbol{\beta}, \boldsymbol{\gamma})}$:

\begin{equation}
    \Braket{H_{C}}_{\boldsymbol{\beta}, \boldsymbol{\gamma}} = \bra{\psi(\boldsymbol{\beta}, \boldsymbol{\gamma})}H_{C}\ket{\psi(\boldsymbol{\beta}, \boldsymbol{\gamma})}.
\end{equation}

\noindent
In general, higher values of $p$ can result in better approximations of the optimal solution, but at the cost of more computations~\cite{Farhi2014}.
For the sake of simplicity, we shall write $\Braket{H_{C}}_{\beta_{1}, \gamma_{1}} = \Braket{H_{C}}_{(\beta_{1}), (\gamma_{1})}$ for $p=1$.

A commonly studied problem for QAOA is the MaxCut problem~\cite{Farhi2014,Truger2022,Li2021,Xue2021,Harrigan2021,Alam2019}. 
It occurs in many application fields, e.g., solid-state physics and integrated circuit design~\cite{Barahona1988} as well as data clustering~\cite{Poland2006}.
Thus, solving the MaxCut problem efficiently helps speeding up computing solutions of these problems.
A \textit{cut} of a graph $G = (V, E)$ splits its set of vertices $V$ into two partitions and can be represented as a bitstring $\boldsymbol{z}\in\{0, 1\}^{|V|}$, where each bit assigns one of the vertices to one of the two partitions.
The size of the cut is the number of edges crossing these two partitions.
A \textit{maximum cut} is at least as large as any other cut of the graph. 
Finding such a cut for a given graph is known as the MaxCut problem, a well-known NP-hard problem~\cite{Karp1972}.

To solve the MaxCut problem with QAOA~\cite{Farhi2014}, the cost Hamiltonian for a graph G is defined as

\begin{equation}
    H_{C_{\text{MaxCut}}} = \frac{1}{2}\sum_{(v,w) \in E}(I - Z_{v}Z_{w})
\end{equation}

\noindent
where $Z_{v}$ is the Pauli Z matrix applied to the $v$-th qubit.
Thus, each individual sum-term $I - Z_{v}Z_{w}$ of $H_{C_{\text{MaxCut}}}$ increases the eigenvalue of a solution $\boldsymbol{z}$ if and only if $z_{v} \ne z_{w}$, that is, if and only if the corresponding edge $(v,w)$ crosses the induced partitions of the cut.
The corresponding unitary operator for the Hamiltonian $H_{C_{\text{MaxCut}}}$ is 
\begin{equation}
    U(H_{C_{\text{MaxCut}}}, \gamma_{l}) = \exp{(-i\gamma_{l}|E|/2)}\prod_{(v,w) \in E}\exp{(i\gamma_{l}Z_{v}Z_{w}/2)}
\end{equation}
which can  be implemented as a product of $R_{ZZ}(-\gamma_{l}) = \exp{(i\gamma_{l}Z_{v}Z_{w}/2)}$ gates up to a global factor $\exp{(-i\gamma_{l}|E|/2)}$.
The corresponding unitary operator for the Hamiltonian $H_{B}$ is 
\begin{equation}
    U(H_{B}, \beta_{l}) = \prod_{v \in V} \exp{(-i\beta_{l}X_{v})}
\end{equation}
which is a product of $R_{X}(2\beta_{l}) = \exp{(-i\beta_{l}X_{v})}$ gates.
The $R_{ZZ}(\gamma_{l})$ gate implements a two-qubit rotation about $Z\otimes Z$ with angle $\gamma_{l}$ and the $R_{X}(2\beta_{l})$ gate implements a single qubit rotation about X with angle $2\beta_{l}$.
Thus, the problem instance of the MaxCut problem determines the ansatz structure since each edge of the graph translates to a two-qubit $R_{ZZ}$ gate in the ansatz.
However, problem structures of practical interest often cannot be trivially mapped to a planar architecture of current NISQ devices with restricted qubit connectivity~\cite{Harrigan2021}. 
Instead, their quantum circuits have to be adapted by inserting additional SWAP operations to permute qubits and adapt the circuit to the connectivity of the quantum device~\cite{Sivarajah2020}.

\subsection{Quantum circuit cutting with gate cuts}

Quantum circuit cutting is a set of techniques that enables to split a quantum circuit into multiple smaller circuits with fewer qubits and gates such that the result of executing the collection of the smaller circuits is the same as the result of executing the original circuit by exploiting subsequent classical postprocessing~\cite{Peng2019,Mitarai2021}.
The primary focus of quantum circuit cutting is on reducing the number of required qubits, but the subcircuits also consist of fewer gates such that cutting may also reduce the circuit depth of the executed circuits.
\begin{figure}
    \centering
    \includegraphics[width=\linewidth]{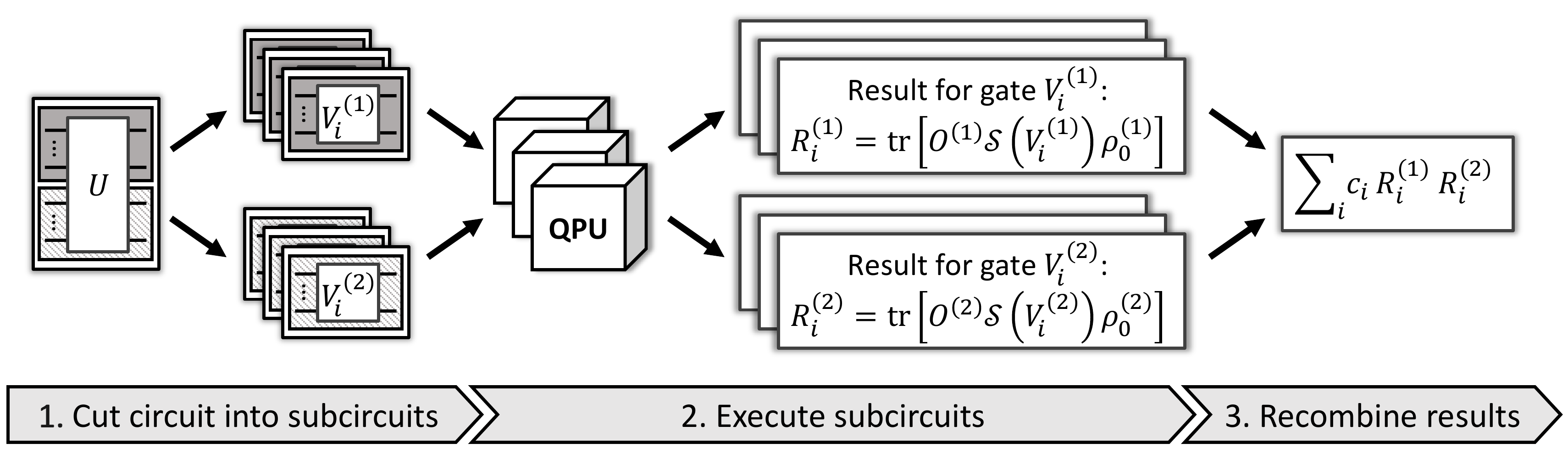}
    \caption{Circuit cutting process: A large quantum circuit gets split into smaller subcircuits by replacing non-local with local operations. The smaller subcircuits get executed on one or more quantum devices and produce the subcircuit results that get recombined to produce the result of the initial circuit.}
    \label{fig:cutting_overview}
\end{figure}
The circuit cutting process consists of three steps: (i)~cutting the circuit into a set of smaller subcircuits, (ii)~execution of these subcircuits, and (iii)~the classical recombination of the subcircuit results.
There are two different approaches to cut a circuit: (i)~\textit{gate cutting} replaces two-qubit gates by sets of local operations~\cite{Mitarai2021,Mitarai2021a} and (ii)~\textit{wire cutting} divides circuit wires carrying quantum information in the form of a qubit into sets of measurement and state-preparation operations~\cite{Peng2019}.
In this work, gate cutting is used, which is presented in \Cref{fig:cutting_overview} and described in more detail below.

Let $\mathcal{H}^{(1)}\otimes \mathcal{H}^{(2)}$ denote a bipartite $n$-qubit quantum system consisting of Hilbert spaces $\mathcal{H}^{(1)}$ and $\mathcal{H}^{(2)}$. The quantum state of this system can be described by a \textit{density operator} $\rho$, which is a positive Hermitian matrix of size $2^{n} \times 2^{n}$ with trace equal to one.
Moreover, consider an $n$-qubit gate represented by a unitary $U$. 
The evolution of the quantum state under the action of gate $U$ can be described by a \textit{superoperator} $\superop(U)$, which is a linear operator that acts on the space of density operators. 
The new state resulting from the application of $U$ on the state described by $\rho$ is given by $\superop(U)\rho = U\rho U^{\dagger}$.

A gate cut of the unitary operator $U$ is its \textit{quasiprobability decomposition~(QPD)} into a set of operators $V_{i}=(V_{i}^{(1)}\otimes V_{i}^{(2)})$ with associated complex factors $c_{i}$ given by $\{(V_{i}, c_{i})\}$ such that  
\begin{equation}
    \superop(U) = \sum_{i}c_{i}\superop\left(V_{i}\right) = \sum_{i}c_{i} \superop\left(V_{i}^{(1)}\right)\otimes \superop\left(V_{i}^{(2)}\right)
\end{equation}
where $V_{i}^{(1)}$ and $V_{i}^{(2)}$ are operators acting on subsystem $\mathcal{H}^{(1)}$ and $\mathcal{H}^{(2)}$, respectively~\cite{Mitarai2021}.
Each operator $V_{i}^{(j)}$ has to be \textit{physically realizable}, that is, its superoperator $\superop(V_{i}^{(j)})$ is a completely positive  linear map that is trace-nonincreasing, i.e., $0 \le \tr[\superop(V_{i}^{(j)})\rho] \le 1$ for any density operator $\rho$~\cite{Nielsen2009}.
Therefore, this includes non-unitary transformations such as projections.
The decomposition of the gate $U$ is depicted in the first step of \Cref{fig:cutting_overview}.

Consider for the partition of the quantum system $\mathcal{H}^{(1)}\otimes \mathcal{H}^{(2)}$ an observable $O=O^{(1)}\otimes O^{(2)}$ that acts independently on each subsystem, and a separable initial state $\rho_{0}=\rho^{(1)}_{0}\otimes\rho^{(2)}_{0}$.
Thus, the subsystems are independent of each other concerning their initialization and measurement.
Then, the evolution of density operator $\rho_{0}$ according to the unitary $U$ and subsequent measurement with observable $O$ can now be reproduced by applying the operations $V_{i}$ and weighting their measurement with $O$ according to $c_{i}$:  
\begin{align}
    \tr\left[O\superop(U)\rho_{0}\right] &= \sum_{i}c_{i} \tr\left[O\superop(V_{i})\rho_{0}\right] \\
    &= \sum_{i}c_{i} \underbrace{\tr\left[O^{(1)}\superop\left(V_{i}^{(1)}\right)\rho^{(1)}_{0}\right]}_{:= R_{i}^{(1)}} \underbrace{\tr\left[O^{(2)}\superop\left(V_{i}^{(2)}\right)\rho^{(2)}_{0}\right]}_{:= R_{i}^{(2)}} \label{eq:posprocessing}
\end{align}
This allows us to evaluate each of the expectation values $R_{i}^{(1)}$ and $R_{i}^{(2)}$ individually and then recombine them to produce the expectation value of the original circuit as shown in step two and three of \Cref{fig:cutting_overview}.
The computational overhead of this procedure can be measured by the number of additional shots needed to approximate the expectation value of the original circuit~\cite{Piveteau2022}. 
Although the expectation value of the original circuit, computed from the subcircuit results, remains unchanged, additional shots are necessary due to the increased variance resulting from sampling the subcircuits. 
This increase in variance is characterized by the factor $\kappa := \sum_{i}|c_i|$ with $\kappa \geq 1$.
Previous studies have demonstrated that achieving a fixed statistical accuracy requires an additional $\mathcal{O}(\kappa^2/\epsilon^2)$ shots for estimating the original expectation value within the error $\epsilon$~\cite{Temme2017}.
It is worth noting that the computational cost of the classical recombination step is limited by the number of shots on the subcircuits, as only the sampled results need to be recombined through multiplication.

Given cuts for gates $U$ and $\tilde{U}$ with QPDs $\{(V_{i}, c_{i})\}$ and $\{(\tilde{V_{i}}, \tilde{c}_{i})\}$, respectively, the cut of their product $U\tilde{U}$ can be constructed as following
\begin{align}
    \superop(U)\superop(\tilde{U}) &= \sum_{i} c_{i} \superop(V_{i}) \sum_{j}\tilde{c}_{j}\superop(\tilde{V_{j}})\\
     &= \sum_{i,j} c_{i}\tilde{c}_{j} \superop(V_{i})\superop(\tilde{V_{j}}).
\end{align}
Thus, a quantum circuit constructed of multiple subsequent gates can be cut by cutting its individual gates.
However, the overhead, measured in terms of the number of additional shots, grows in the worst case exponentially as each cut introduces a multiplicative factor $\kappa$~\cite{Piveteau2022}.

\subsection{Gate cutting for QAOA's MaxCut ansatz}\label{sec:cutting_rzz}
The only multi-qubit gates in the MaxCut ansatz for QAOA are $R_{ZZ}$ gates.
Thus, the ansatz for a given graph can be cut by partitioning its qubits into separate parts and then cutting all $R_{ZZ}$ gates between these qubit partitions.
To cut a single $R_{ZZ}$ gate, we use the QPD for cutting arbitrary two-qubit gates introduced by Mitarai and Fujii~\cite{Mitarai2021,Mitarai2021a}:

\begin{equation}\label{eq:rzz_cut}
    \begin{split}
    \superop\!\left(R_{ZZ}(\gamma)\right)\! &=  \cos^{2}\left(\frac{\gamma}{2}\right) \superop\!\left(I\otimes I\right)+ \sin^{2}\left(\frac{\gamma}{2}\right) \superop\!\left(Z\otimes Z\right)
    \\&\quad+ \cos\left(\frac{\gamma}{2}\right)\sin\left(\frac{\gamma}{2}\right)\left(A\otimes B + B\otimes A\right)
\end{split}
\end{equation}

\noindent
with 

\begin{equation}\label{eq:A}
    A := \superop\!\left(R_Z\!\left(-\frac{\pi}{2}\right)\!\right) - \superop\!\left(R_Z\left(\frac{\pi}{2}\right)\!\right)
\end{equation}

\noindent
and

\begin{equation}
    B := \superop\!\left(\dfrac{I - Z}{2}\right) - \superop\!\left(\dfrac{I + Z}{2}\right).
\end{equation}

\noindent
The two operations forming $B$ are the projections on the states $\ket{1}$ and $\ket{0}$, respectively, i.e., $(I-Z)/2 = \ketbra{1}{1}$ and $(I+Z)/2 = \ketbra{0}{0}$.
Although these operations are not unitary and thus cannot be directly implemented as a gate of a quantum circuit, they can be realized by a post-selective measurement~\cite{Mitarai2021}.
By applying this QPD, the circuit with the $R_{ZZ}$ gate can be replaced by circuits that perform at each of the two previously connected qubits one of the following five operations instead: an $I$, $Z$, $R_Z\left(-\frac{\pi}{2}\right)$, or $R_Z\left(\frac{\pi}{2}\right)$ gate, or a measurement for the projections.
Since the two qubits are now separable, ten different subcircuits consisting of the five operations for the upper and five for the lower qubit must be executed.
\ref{ap:rzz_cut} provides a proof of the cutting formula in \Cref{eq:rzz_cut}.

\section{Related Work}\label{sec:relatedwork}

An umbrella term that combines techniques to perform the computation of a large quantum circuit as the execution of smaller subcircuits is \textit{circuit knitting}~\cite{Bravyi2022}.
It includes circuit cutting and \textit{entanglement forging}~\cite{Eddins2021,Huembeli2022}, which is a closely related technique.

In circuit cutting experiments, various works have practically demonstrated that circuit cutting can be used to expand the size of executable circuits beyond the physical capabilities of NISQ devices, e.g., for a set of benchmark circuits~\cite{Tang2021}, GHZ circuits~\cite{Ayral2020}, and circuits for linear cluster states~\cite{Ying2022}.
Moreover, it has been shown experimentally that in noisy simulations and on NISQ devices, circuit cutting can lead to better, i.e. less noisy, execution results~\cite{Tang2021,Ying2022,Ayral2020,Ayral2021}.
Circuit cutting can help with gate errors and short decoherence times since smaller circuits are less susceptible to these noise sources; in contrast, readout errors can be detrimental to the circuit cutting procedure since it requires additional measurements~\cite{Ayral2021}.
In addition, it was investigated how errors occurring in different subcircuits affect the final recombined result, and thereby, different error propagation characteristics and sensitivities were found for different subcircuits~\cite{Casciola2022}.
\looseness=-1

The cost of circuit cutting grows in the worst case exponentially with the number of cuts~\cite{Lowe2022}.
Since the introduction of the first gate-cutting technique~\cite{Bravyi2016}, various methods and adaptions that lower the cost have been developed~\cite{Peng2019,Mitarai2021,Mitarai2021a,Piveteau2022,Lowe2022}.
Peng \etal\cite{Peng2019} present the first wire cut, which cost is reduced by Lowe \etal\cite{Lowe2022} via randomized measurements.
Mitarai and Fujii~\cite{Mitarai2021,Mitarai2021a} lower the cost of a gate cut.
Moreover, based on gate teleportation~\cite{Gottesman1999}, Piveteau and Sutter~\cite{Piveteau2022} show that extending gate-cutting with classical communication between subcircuits can reduce the overhead further, although the overhead is still exponential.

In addition, further improvements and tooling around the cutting methods themselves were developed.
A cutting framework that automatically determines optimal cut locations and quickly finds solution states of large quantum circuits was introduced~\cite{Tang2021,Tang2022}. 
This framework is part of Qiskit's Circuit Knitting Toolbox, which is a collection of tools to decompose quantum circuits~\cite{circuit_knitting_toolbox}.
Furthermore, maximum-likelihood tomography applied to wire cutting produces better results by ensuring non-negativity and normalization for the computed result distribution~\cite{Perlin2021}.
This method estimates the result of a circuit even in noise-free experiments with higher fidelity than the execution of the full circuit.
Moreover, to lower the number of subcircuits, a machine learning approach applied to gate-cutting can approximate the result of the original circuit with a significantly reduced set of subcircuits~\cite{Marshall2022}.

Furthermore, there is work that investigates partitioning circuits of VQAs.
Saleem \etal\cite{Saleem2021} apply circuit cutting in a VQA to solve the Maximum Independent Set problem.
However, they execute their approach only on a simulator since the complex multi-control gates used result in   circuits that are too noise sensitive for current NISQ devices.
Moreover, Lowe \etal\cite{Lowe2022} apply wire cuts in QAOA and perform numerical experiments on a simulator to show the speed up of their method.
Both works do not investigate the effect of circuit cutting on VQAs running on error-prone NISQ devices.
Additionally, it has been demonstrated for a classification task that a restricted ansatz consisting of multiple local circuit partitions can be used to improve the training of a VQA and its robustness against noise~\cite{Tueysuez2022}.
However, this approach does not allow any interaction between these partitions, and therefore, this technique cannot generally implement the exact ansatz defined by QAOA's cost Hamiltonian of a specific problem instance.

Besides methods on the circuit level, decomposition approaches on the algorithmic level exist~\cite{Li2021,Zhou2022,
Shaydulin2019,Shaydulin2019a,Tomesh2021,UshijimaMwesigwa2021,Ayanzadeh2022}.
Examples include divide and conquer approaches for QAOA~\cite{Li2021,Zhou2022} and quantum local search methods that iteratively optimize local subproblems of a problem~\cite{Shaydulin2019,Shaydulin2019a,Tomesh2021,UshijimaMwesigwa2021}.
In addition, Ayanzadeh \etal\cite{Ayanzadeh2022} demonstrate for the MaxCut problem that by removing nodes with high connectivity from the graph and accounting for their possible assignments in the objective function, the circuit size can be reduced, and the fidelity of QAOA on NISQ devices can increase.

\section{Motivation and research questions}\label{sec:motivation}
Due to the already discussed problems of NISQ devices, algorithms for applications on them are an active research field.
One promising pathway is the ongoing development of VQAs~\cite{Cerezo2021}, like the prominent QAOA.
Since the introduction of QAOA~\cite{Farhi2014}, much work has been done to develop the algorithm further to achieve quantum advantage on NISQ hardware.
For example, new objective functions have been introduced \cite{Li2020,Barkoutsos2020}, research on warm-starting QAOA has been done~\cite{Egger2021,Truger2022}, and different modifications of QAOA have been presented~\cite{Baertschi2020,Bravyi2020,Herrman2022}.

However, noise is still an open problem for QAOA. 
It limits the advantage of increasing $p$ to improve QAOA solutions in practical experiments on NISQ devices~\cite{Alam2019}.
This issue is particularly pronounced for problem instances that have connectivity that does not match the hardware, as the quality of results rapidly declines with increasing problem size due to the added noise from SWAP operations~\cite{Harrigan2021}.
In general, noise in the computation of VQAs leads to exponentially vanishing gradients in the training landscape~\cite{Xue2021,Franca2021}. 
This effect is referred to as \textit{noise-induced barren plateaus~(NIBPs)}~\cite{Wang2021}.
Although NIBPs and \textit{noise-free barren plateaus}~\cite{McClean2018,Cerezo2021a} both lead to vanishing gradients, they differ in the way they affect the training landscape.
Noise-free barren plateaus allow the global minimum to reside inside a deep, narrow valley, while NIBPs exponentially flatten the entire landscape~\cite{Franca2021,Wang2021}.
Notwithstanding the fact that noise does not change the optimization direction in the QAOA parameter space, and optimal parameters stay nearly the same~\cite{Xue2021}, resolving the exponential concentrated training landscapes to a fixed precision requires an exponential number of shots~\cite{Wang2021,Wang2021a}.
Moreover, although gradient-free optimization methods do not use gradient information, they also scale exponentially in the number of shots for ansätze with exponentially small gradients~\cite{Arrasmith2021}.

Since common strategies that address noise-free barren plateaus do not work for NIBPs, two basic strategies for preventing NIBPs remain~\cite{Wang2021}: (i)~lowering the hardware noise level and (ii)~reducing the circuit complexity, mainly focusing on the depth but also the width of the circuit.
While error mitigation can reduce the effect of hardware noise in VQAs~\cite{Barron2020,Beisel2022}, a broad class of these techniques cannot remove the connected exponential resource scaling~\cite{Wang2021a}.

In this work, we apply quantum circuit cutting to reduce the complexity of the executed circuits.
Quantum circuit cutting can be classified as a decomposition technique on the quantum circuit level as it is independent of the algorithm generating the circuits~\cite{Peng2019,Mitarai2021}.
Like other decomposition approaches, it can reduce the number of required qubits and increase the size of solvable problem instances with available quantum devices~\cite{Tang2021,Ying2022}.
However, our work focuses not on expanding the problem size but on decreasing circuit complexity.
Hence, we aim to decrease the susceptibility of noise during computation and consequently increase the result quality with circuit cutting.
Such an examination of circuit cutting on NISQ devices in the context of a VQA, e.g., QAOA, is missing. 
In particular, there is a lack of research on the extent to which circuit cutting applied in QAOA can help to reduce the influence of noise and whether its application can lead to better results. 
Thus, to address this issue, we formulate the main \textit{research question~(RQ)} of this work:

\researchquestion{To what extent can circuit cutting improve the results of QAOA when executing on NISQ devices?}

To answer our main RQ, we refine it into two \textit{sub-RQs~(SRQs)} that need to be answered.
First, we focus on what impact cutting the QAOA ansatz has on the objective function.
Of particular interest is whether the reduced size of subcircuits can help to reduce the effect of NIBPs.
The first SRQ is as follows:

\subresearchquestion{How does cutting the QAOA ansatz influence its corresponding objective function on NISQ devices?}

The second part includes the classical optimization of the parameters for the cut ansatz.
We investigate how cutting influences the classical optimization process in QAOA.
In particular, we are interested in whether cutting helps to obtain better solutions on NISQ devices.
This is formulated in our second SRQ:

\subresearchquestion{Can QAOA with circuit cutting obtain better solutions on NISQ devices when the entire algorithm, including parameter optimization, is executed?}

\section{Research design}\label{sec:design}
To tackle these RQs, we design and conduct two experiments focussing on the MaxCut problem for unweighted graphs.
In the following, we introduce our adaptions to the gate cut described in \Cref{sec:cutting_rzz} and afterward give a general overview of the experimental setup that all experiments have in common.
Subsequently, we describe in detail the design for each experiment.
In the end, we introduce metrics to evaluate the obtained results.
\looseness=-1

\subsection{Adaptions to the cut of the $R_{ZZ}$ gate}
We adapt the approach described in \Cref{sec:cutting_rzz} to reduce the number of subcircuits.
Our adjustment of \Cref{eq:A} is as follows:
\begin{equation}
    A = \superop(I) + \superop(Z) - 2 \superop\!\left(R_Z\left(\frac{\pi}{2}\right)\!\right).
\end{equation}
The mathematical details are provided in \ref{ap:rz_decomposition}.
By harnessing this equality, the number of different subcircuits of the $R_{ZZ}$ gate decreases from ten to eight since the two local $R_Z\left(-\frac{\pi}{2}\right)$ gates disappear and $A$ can be expressed as a weighted sum of the $I$, $Z$, and $R_Z\left(\frac{\pi}{2}\right)$ gates.
The results for the  $I$ and $Z$ operations can be reused as they are stored in classical memory.
Although this substitution reduces the number of subcircuits, it introduces larger factors $c_{i}$ in the QPD.
This does not change the equality, but it leads to a higher variance, and therefore, more shots are needed for convergence~\cite{Piveteau2022a}.
However, this can be a worthwhile trade-off for a small number of cuts as the factors for them remain relatively small. 
This technique is applied in all our experiments.

\subsection{Overview of the experimental setup}

This section describes the steps that all experiments have in common.
At a high level, each experiment consists of three parts as depicted in \Cref{fig:exp_design_generation}: (i)~the generation of the problem instance, the corresponding circuits, and subcircuits, (ii)~the execution of the latter on quantum devices, and (iii)~the postprocessing including the recombination of subcircuit results and the evaluation of the objective function.
Experiments can repeat steps (ii) and (iii) in a variational optimization loop to refine the circuit parameters with a classical optimizer until a termination condition is met.
Hereafter, we will call the QAOA ansatz that uses circuit cutting \textit{cut-QAOA}.
As quantum devices for the execution, we use IBM's superconducting hardware.
The experiments are implemented with \textit{Qiskit}~\cite{Qiskit}.
Wherever a classical optimizer is necessary, COBYLA, in its default configuration from Qiskit, is employed.
Since the optimizer only supports minimization, we transformed the maximization of QAOA's expectation value in our implementation into a minimization problem by flipping the sign of the classically computed objective function. 
Thus, the optimal value corresponds to the minimal objective value.
In the following, the individual steps are described in more detail.

\subsubsection{Problem instance generation, circuit generation, and cutting.}
\begin{figure}
    \centering
    \includegraphics[width=\linewidth]{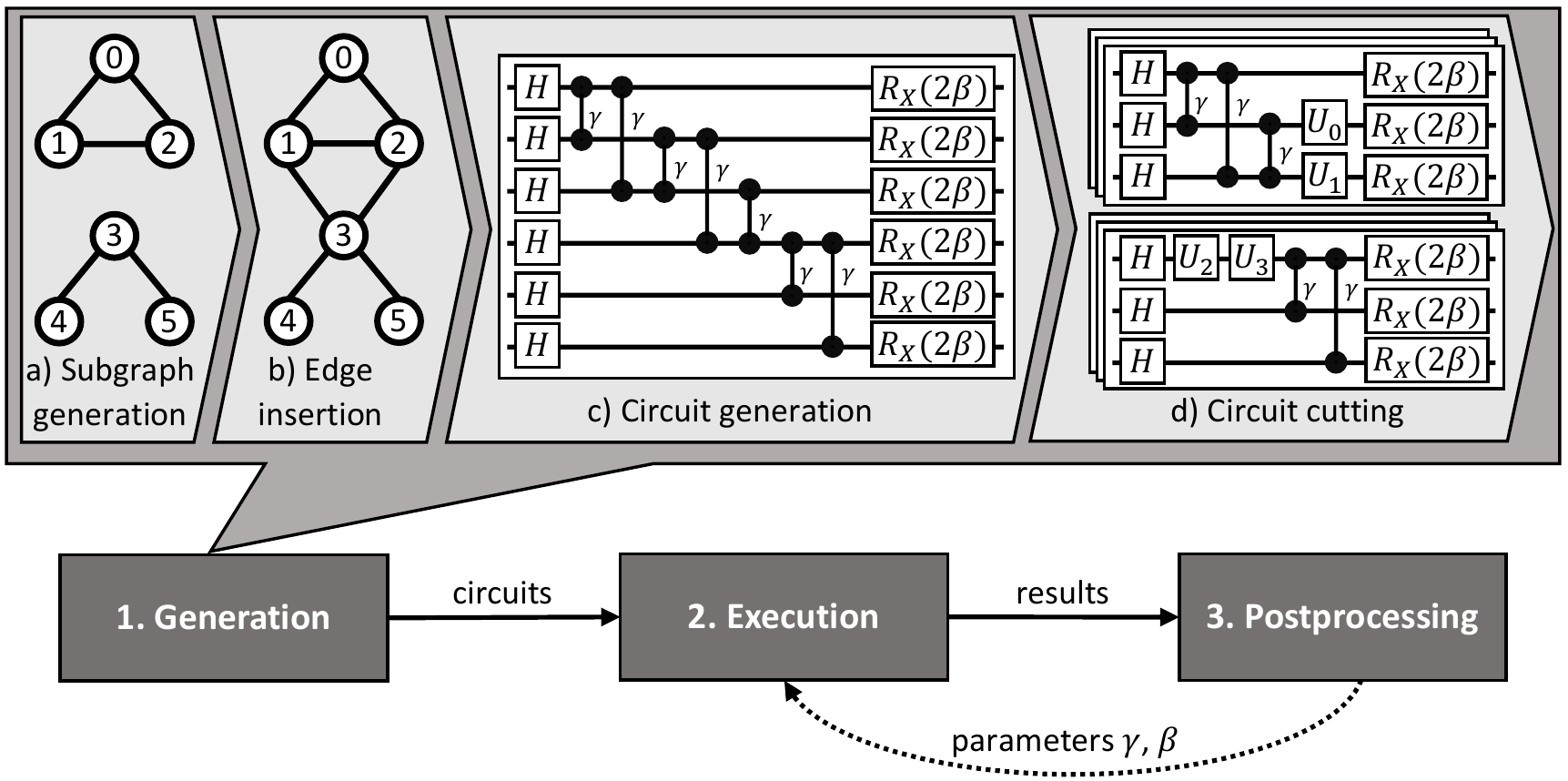}
    \caption{General overview of the evaluation process of the experiments with a refined preprocessing example consisting of problem generation, circuit generation, and circuit cutting.}
    \label{fig:exp_design_generation}
\end{figure}

First, a problem instance has to be created.
This is a crucial step since the edges of the graph map to the $R_{ZZ}$ gates in the QAOA ansatz and therefore, the graph structure determines the number of cuts that are later needed to separate the ansatz.
To enable the resulting ansatz to be separable with a specific number of cuts, we first generated two independent connected random subgraphs of equal size with the \textit{Erdős–Rényi} $G(n,p)$ model~\cite{erdHos1960evolution}, i.e., each possible edge in a graph with $n$ vertices is selected with the probability $p$.
Here, we enforce that each subgraph is connected by repeating the generation procedure until it yields two connected subgraphs.
This is illustrated in step \textit{a} of the generation in \Cref{fig:exp_design_generation}.
Afterward, we connect the two subgraphs by randomly inserting two edges between them and thereby fixing the locations of two intended cuts (see step \textit{b}).

Following this, we generate an ansatz based on the resulting graph where the parameters can be inserted afterward (see step \textit{c}).
In our experiments, only QAOA ansätze with depth $p=1$ are considered, meaning that the cost Hamiltonian and mixer Hamiltonian are executed only once.
Although the IBM quantum devices used in our experiments do not natively support $R_{ZZ}$ gates, we use these gates in the circuit generation phase to apply our cutting procedure and leave it to the transpiler to replace the remaining $R_{ZZ}$ gates with native gates, i.e., two CNOT gates and a $R_{Z}$ rotation.

\begin{figure}
    \centering
    \includegraphics[width=\linewidth]{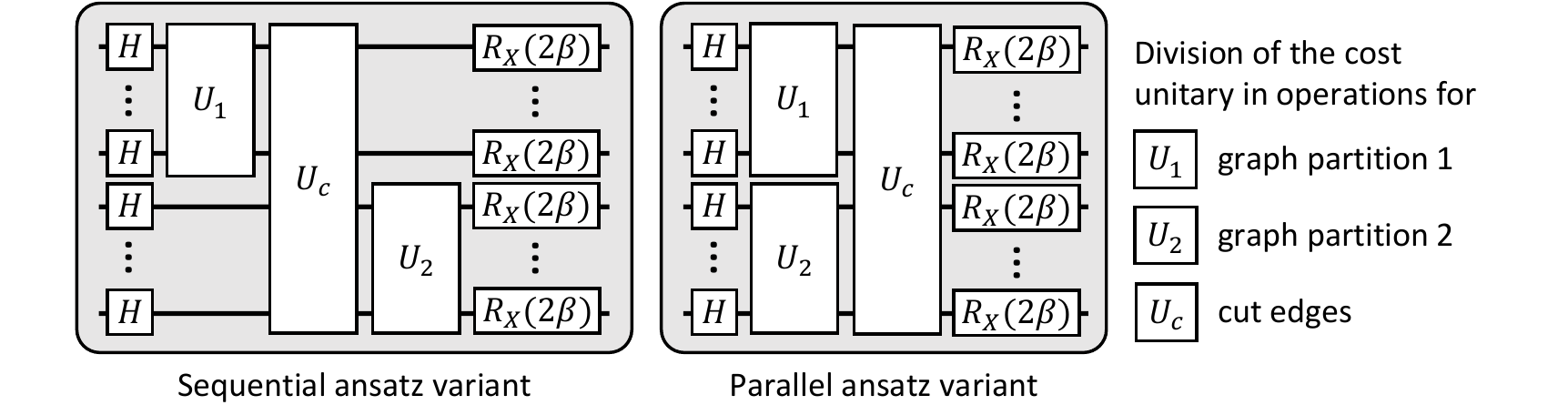}
    \caption{Schematic representation of both QAOA ansatz variants.}
    \label{fig:exp_qaoa_compilation}
\end{figure}

There are multiple equivalent QAOA ansätze for each problem instance since the $R_{ZZ}$ gates in the cost unitary $U(H_{C}, \gamma_{l})$ commute, and thus they can be arranged in an arbitrary order.
Different orderings of the $R_{ZZ}$ gates have no impact on a noise-free device and lead to equivalent results.
However, this is different for NISQ devices.
Depending on the arrangement of operations, the depth of the circuit may vary. 
In addition, specific arrangements of operations can be mapped to the restricted connectivity of the quantum device with fewer additional SWAP operations and consequently lead to transpiled circuits of a lower depth.
Moreover, independent simultaneous operations on different qubits may affect each other unintendedly and hence are another source of noise~\cite{AshSaki2020}.
All these issues have to be taken into account for optimal circuit transpilation.
However, optimal circuit transpilation is NP-complete, and transpilers rely mostly on heuristics~\cite{Paler2021}.
Moreover, the used Qiskit transpiler does not conduct all these optimizations without providing custom transpiler passes~\cite{qiskit_transpiler_passes}.
Therefore, to incorporate the effect of different arrangements of operations in our experiments and compare them to the effect of circuit cutting,
we consider this in our experiments manually and generate two equivalent ansätze for each problem instance.
One ansatz optimizes the number of simultaneous gates in the circuit by executing the gates of the two graph partitions sequentially, and the other ansatz optimizes the depth of the circuit by executing the gates of the two graph partitions in parallel.
They are schematically depicted in \Cref{fig:exp_qaoa_compilation}.
The sequential version of the ansatz starts with all $R_{ZZ}$ operations of the first partition. 
Then we apply the operations that get cut. 
Finally, we add the $R_{ZZ}$ operations for the second partition.
The parallel version of the ansatz applies all $R_{ZZ}$ operations of both graph partitions in parallel and then the gates that get cut in a final step. 
We use the same arrangement of $R_{ZZ}$ operations within the partitions in both versions as well as in the subcircuits of the cut ansatz.
The sequential version results in a deeper circuit with fewer parallel operations. 
In contrast, the parallel ansatz is shallower but with more parallel operations.

Next, the generated circuit gets cut.
To produce its subcircuits that are depicted in step \textit{d} of \Cref{fig:exp_design_generation}, we use the following procedure:
\begin{enumerate}
    \item Remove all multi-qubit operations between the different qubit partitions corresponding to the two subgraphs such that the circuit decomposes into separable circuit fragments.
    \item Generate for each of these circuit fragments their respective subcircuits by inserting all possible combinations of $I$, $Z$, $R_Z\left(\frac{\pi}{2}\right)$ and measurement at the positions of the removed gates.
\end{enumerate}

\subsubsection{Execution.}
The execution includes not only running the subcircuits for cut-QAOA but also the original QAOA circuits, which will be used later in the evaluation for comparison purposes.
But before an experiment's generated circuits and subcircuits can be executed, their parameters must be fixed.
The parameters are defined either as hyperparameters of the experiment or are selected by the classical optimizer as part of the variational loop.
Afterward, we transpile all circuits for the selected quantum device with Qiskit's internal transpiler in its default configuration~\cite{Qiskit}.
All transpiled circuits get batched into jobs and submitted jointly to the quantum device for execution.
In our experiments, we executed the original QAOA ansatz with the same number of shots as the total amount of shots across all subcircuits of cut-QAOA. 
Consequently, the same number of shots were spent with and without cutting in our experiments.
For the different subcircuits of a cut-QAOA ansatz, we evenly distribute the number of shots among them.
Another possibility would be to distribute the shots across the subcircuits in proportion to their factor $c_{i}$ in the QPD.
However, we choose not to do so since Qiskit does not allow to specify the number of shots per circuit but only for all circuits in one job.

\subsubsection{Postprocessing.}
Lastly, postprocessing is performed.
Thereby, the results of all subcircuits get processed and recombined according to \Cref{eq:posprocessing} to reproduce the result of the original circuit.
Subsequently, the MaxCut objective value is calculated for all results, i.e., of the original circuits' execution results and the subcircuits' recombined results.

\subsection{Description of the experiments}
In the following, a detailed description of the two experimental setups used for the evaluation is given.

\subsubsection{Experiment 1: Objective function.}
To investigate RQ~1, we perform an experiment that evaluates the function at equally distributed parameter configurations.
More precisely, it consists of the following steps:
\begin{enumerate}
    \item \textbf{Generate problem instance and corresponding circuits:} 
    For this experiment, we randomly generated problem instances with the method described above that lead to separable circuits with two cuts.
    We use the two introduced variations of the ansatz.
    \item \textbf{Cut circuit into subcircuits:} 
    Although we generate and execute the sequential and parallel ansatz for later comparison, we only cut one of the two ansätze to keep the number of circuits to be executed smaller, allowing us to execute more runs of the experiment with different graphs.
    In contrast to the original circuits, the subcircuits of both ansätze have the same depths and the same number of parallel operations since they  have the same arrangement of two-qubit gates by design and differ only in the position of the cut.
    Therefore, we choose to cut only the sequential variant of the ansatz into the predefined partitions and then generate its parameterized subcircuits.
    \item \textbf{Create circuits with parameters:} 
    Since the parameters define rotations about the angles $2\beta$ and $\gamma$, the objective function is periodic, and we can restrict the parameters to $(\beta, \gamma) \in D=([0, \pi],  [0, 2\pi])$. 
    Therefore, we sample the parameter space $D$ with an equidistant grid 
    with $20 \times 40$ points resulting in 800 function evaluations.
    For each parameter configuration of the grid, the two ansatz variants of the original circuit and the subcircuits with fixed parameters are created.
    \item \textbf{Execute circuits:}
    All circuits and subcircuits are transpiled for and executed on the selected quantum device.
    We extract from the execution not only the aggregated result of all shots but also the results of every single shot.
    This enables us to analyze how the result behaves with different shot numbers by considering only parts of the shots, e.g., the first 1000 shots from execution with 10000 shots.
    \item \textbf{Recombine subcircuits:}
    In classical postprocessing, we recombine the subcircuits with the procedure above.
    \item \textbf{Evaluate objective function:}
    The objective values for the results of the original circuits and the cut circuits are computed for different shot numbers.
\end{enumerate}

\subsubsection{Experiment 2: QAOA execution.}
To investigate RQ~2, we run the entire VQA, including parameter optimization for QAOA and cut-QAOA, ten times on the same problem instance.
In each execution, new randomly selected initial parameters were used in order to reduce the influence of the initial parameters and reason over them in the later analysis.
To allow later evaluation across the two experiments, we conducted this experiment with the same problem instances as the experiment above.
Furthermore, to minimize the effect of different calibrations of the quantum devices, this experiment was always performed directly following the previous experiment for each problem instance.
In detail, this experiment consists of the following steps:
\begin{enumerate}
    \item \textbf{Generate problem instance:}
    We use the randomly generated problem instance from the experiment above.
    \item \textbf{Repeat:} We repeat the following steps ten times.
    \begin{enumerate}
        \item \textbf{Choose initial parameters:} 
        We randomly choose initial parameters $\beta \in [0, \pi]$ and $\gamma \in [0, 2\pi]$. 
        \item \textbf{Execute QAOA:}
        The QAOA is performed twice for the generated problem instance: once for each presented ansatz variant.
        Each run gets started with the chosen initial parameters.
        \item \textbf{Execute cut-QAOA:}
        Subsequently, the cut-QAOA is executed on the same initial parameters.
    \end{enumerate}
\end{enumerate}

\subsection{Metrics for the evaluation of the objective functions}
In this section, we introduce metrics to assess and quantify the objective functions computed on the quantum devices in our first experiment for both QAOA variants, QAOA and cut-QAOA.
Therefore, we consider metrics for the objective function themselves and its gradients.
First, we assess the mean absolute objective value difference between a QAOA variant on a quantum device and QAOA without cutting on the simulator. 
The smaller the difference, the closer the computed objective values of the NISQ device are to the noise-free values of the simulator.
The \textit{mean absolute difference~(MAD)} between the simulated objective function $\Braket{H_{C}}_{\beta, \gamma}^{\text{SIM}}$ and the computation on a NISQ device $\Braket{H_{C}}_{\beta, \gamma}^{\text{QPU}}$ for the finite parameter sets $P_{\beta} \subset [0, \pi] $ and $P_{\gamma} \subset [0, 2\pi]$ is 

\begin{equation}
    \operatorname{MAD}_{\text{SIM}} = \frac{1}{|P_{\beta}||P_{\gamma}|}\sum_{\beta \in P_{\beta}, \gamma \in P_{\gamma} }\left|\Braket{H_{C}}_{\beta, \gamma} ^{\text{QPU}}  - \Braket{H_{C}}_{\beta, \gamma} ^{\text{SIM}}\right|.
\end{equation}

\noindent
Secondly, we have examined the mean absolute objective value difference between a QAOA variant on a NISQ device and the \textit{maximally mixed state~(MMS)}, i.e., the state with density operator $I/2$.
It measures the concentration of the objective function and is an indicator for NIBPs~\cite{Wang2021}.
The larger its distance from the objective value of the MMS, the more the computed objective values of the QAOA variant deviate from the objective value achieved by randomly picking states with equal probability.
The MAD from the objective value of the MMS of $n$ qubits is

\begin{equation}
    \operatorname{MAD}_{\text{MMS}} = \frac{1}{|P_{\beta}||P_{\gamma}|}\sum_{\beta \in P_{\beta}, \gamma \in P_{\gamma} }\left|\Braket{H_{C}}_{\beta, \gamma}  - \frac{1}{2^{n}}\tr\left[H_{C}\right]\right|.
\end{equation}

\noindent
This metric can be calculated for the NISQ device and the simulator.
We compute the objective value of the MMS classically by sampling uniformly at random from the solution space and computing the expected objective value for the sample. 

Furthermore, we assess the gradients of the objective function's computed parameter landscapes as NIBPs lead to vanishing gradients.
The gradient of the objective function is

\begin{equation}
    \nabla \Braket{H_{C}}_{\beta, \gamma} = \left(\frac{\partial}{\partial\beta}\Braket{H_{C}}_{\beta, \gamma}, \frac{\partial}{\partial\gamma}\Braket{H_{C}}_{\beta, \gamma}\right)^{T}.
\end{equation}

\noindent
We numerically computed the partial derivatives using second-order accurate central differences~\cite{Quarteroni2007}.
To assess the gradients of the computed parameter landscapes, we consider their size~\cite{Wang2021} and the mismatch in their direction compared to the simulation.
For the size of the gradient, we have taken into account their average size and the variance in size.
The average gradient size is defined as

\begin{equation}
    M = \frac{1}{|P_{\beta}||P_{\gamma}|}\sum_{\beta \in P_{\beta}, \gamma \in P_{\gamma} } \left\lVert \nabla \Braket{H_{C}}_{\beta, \gamma}  \right\rVert_{1}
\end{equation}

\noindent
where $\left\lVert \cdot \right\rVert_{1}$ is the 1-norm.
The variance of the gradient size is 

\begin{equation}
    \frac{1}{|P_{\beta}||P_{\gamma}|}\sum_{\beta \in P_{\beta}, \gamma \in P_{\gamma} } \left(\left\lVert \nabla \Braket{H_{C}}_{\beta, \gamma}  \right\rVert_{1} - M \right)^{2}.
\end{equation}

\noindent
A decreasing variance means that smaller and smaller changes in the gradients occur, which makes it more and more difficult for a classical optimizer to find an optimum.
Besides the pure size of the gradients, the gradients must also guide the classical optimizer to the optima. 
For this, we analyze the direction of the gradients.
We compare the gradient direction from the NISQ device with the simulated gradients.
To this end, we employ the average pairwise cosine similarity between those gradients~\cite{Ilyas2020,Kim2020}.
It is defined as follows:

\begin{equation}
    \frac{1}{|P_{\beta}||P_{\gamma}|}\sum_{\beta \in P_{\beta}, \gamma \in P_{\gamma} } \frac{\nabla \Braket{H_{C}}_{\beta, \gamma}^{\text{QPU}} \cdot \nabla \Braket{H_{C}}_{\beta, \gamma}^{\text{SIM}}}{\left\lVert \nabla \Braket{H_{C}}_{\beta, \gamma}^{\text{QPU}}\right\rVert_{2} \left\lVert \nabla \Braket{H_{C}}_{\beta, \gamma}^{\text{SIM}}\right\rVert_{2}}.
\end{equation}

\section{Results}\label{sec:results}

This section presents the results for the two introduced experiments in order to answer our RQs.
Since the differences in the results between the sequential and parallel ansatz of the QAOA are minor compared to the results of cut-QAOA and since, in direct comparison, the parallel ansatz performs slightly better on average, we show only the results for the parallel ansatz in this section for the sake of clarity.
However, the code and all data generated during the experiments are publicly available at~\cite{codeanddata}.

\subsection{Results for experiment 1: Objective function}
Following our experiment design, the first question to consider is how cutting the QAOA ansatz influences the corresponding objective function using NISQ devices. 
In the following, this is first discussed in detail, for one example graph, and then the results for the whole dataset are presented.

\begin{figure}
     \centering
     \begin{subfigure}[t]{0.49\linewidth}
         \centering
         \includegraphics[width=\linewidth]{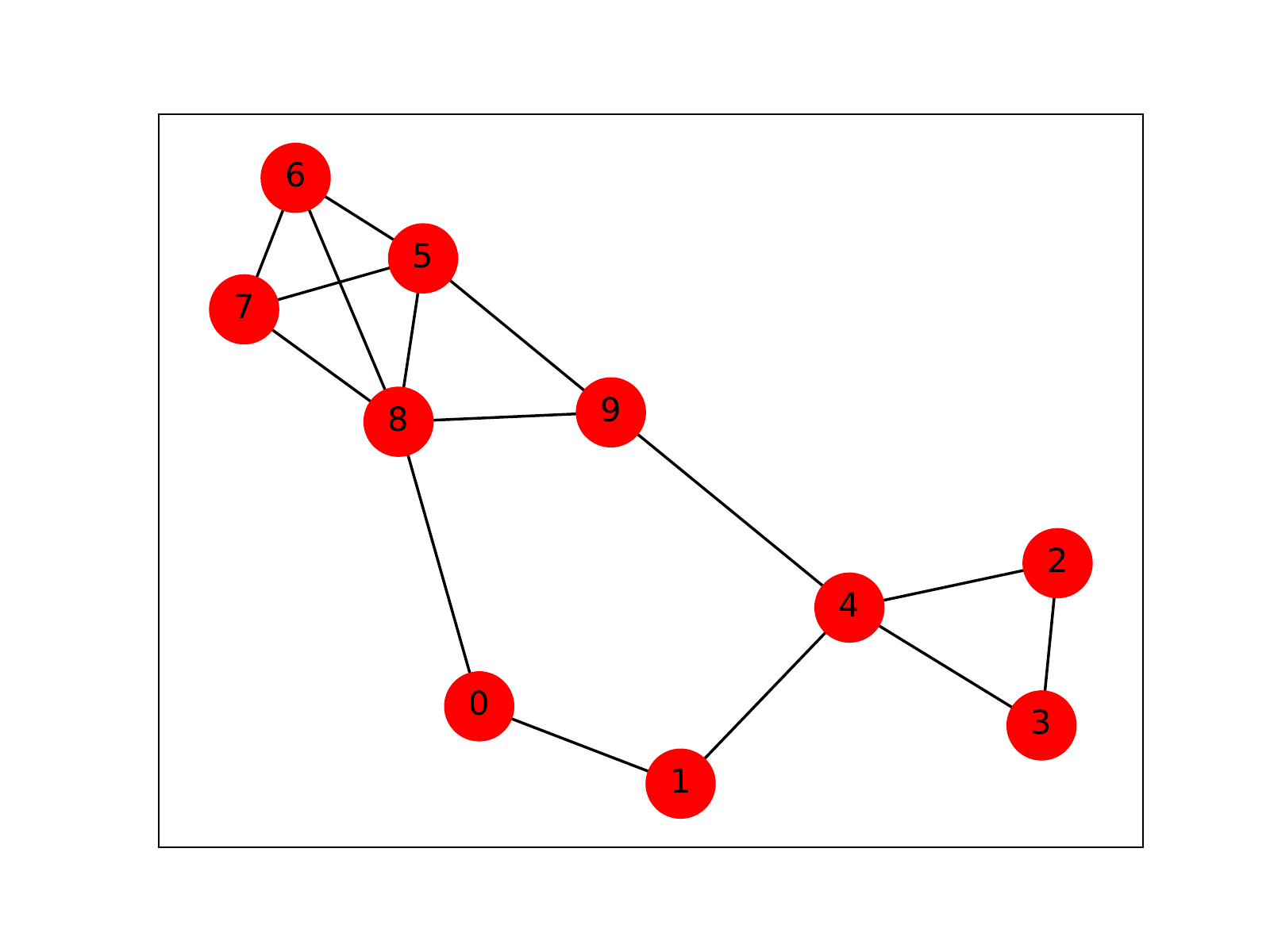}
         \caption{Example graph to apply MaxCut using QAOA.}
         \label{fig:exp1_graph}
     \end{subfigure}
     \hfill
     \begin{subfigure}[t]{0.49\linewidth}
         \centering
         \includegraphics[width=\linewidth]{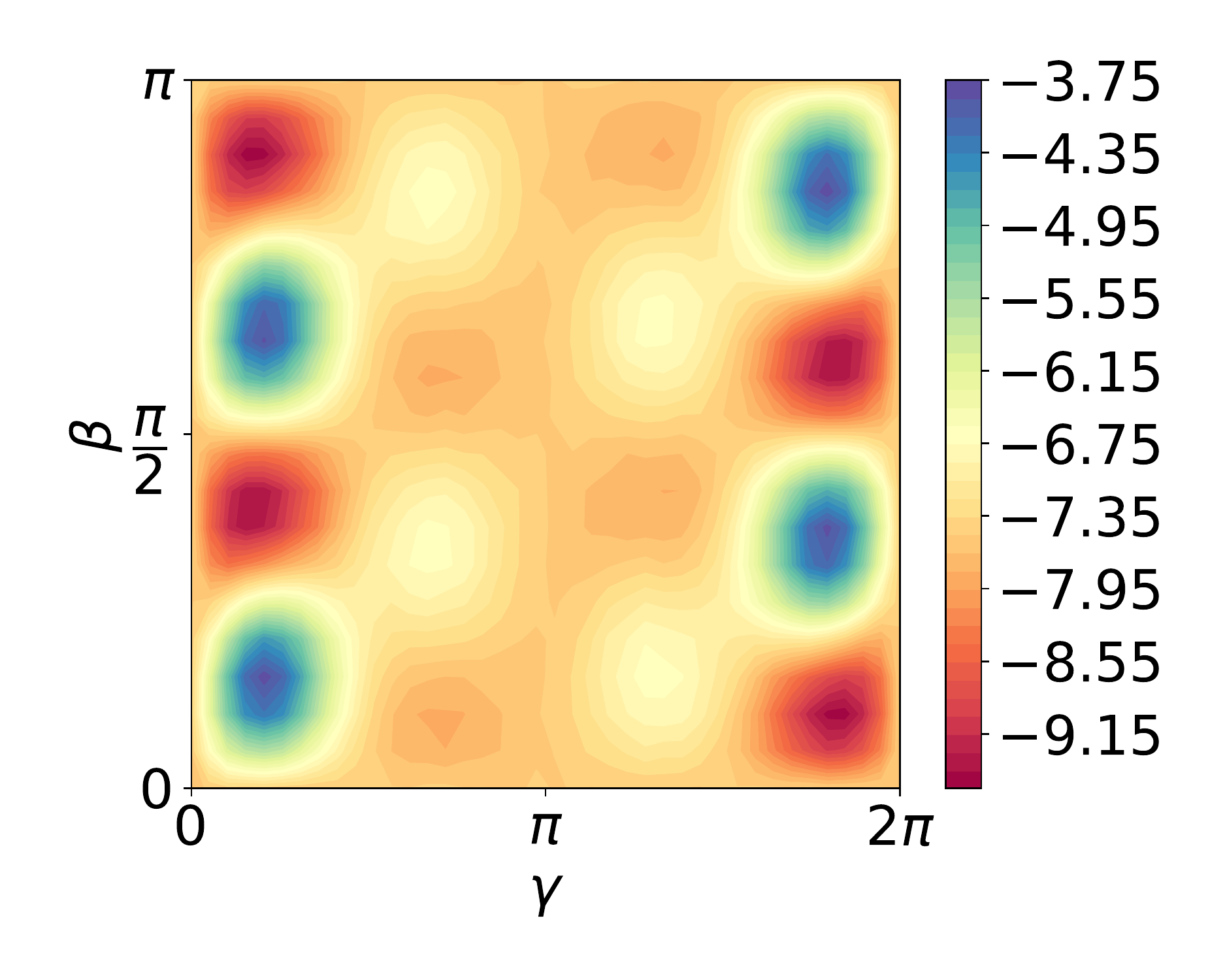}
         \caption{Objective value in the parameter landscape for the example graph simulated for $p=1$.}
         \label{fig:exp1_sim_param_map}
     \end{subfigure}
        \caption{(a) Example graph from the data set and (b) its QAOA objective function for the MaxCut problem in noise-free simulation.}
        \label{fig:exp1_data}
\end{figure}

\subsubsection{Example graph}

In this section, the computed objective functions for an example graph are visualized and evaluated using the introduced metrics for the experiment.
\Cref{fig:exp1_data} shows the ten-node example graph (\Cref{fig:exp1_graph}) with its corresponding noise-free objective function (\Cref{fig:exp1_sim_param_map}).
The cut of edges (0, 8) and (4, 9) separates the graph into two subgraphs of equal size.
The objective function is evaluated on a simulator for $p=1$ QAOA layers.
The desired minima are shown in shades of red in the colorized parameter map.
The higher the objective value, the more the color changes to a blue tone.

\begin{figure}
     \centering
     \includegraphics[width=0.8\linewidth]{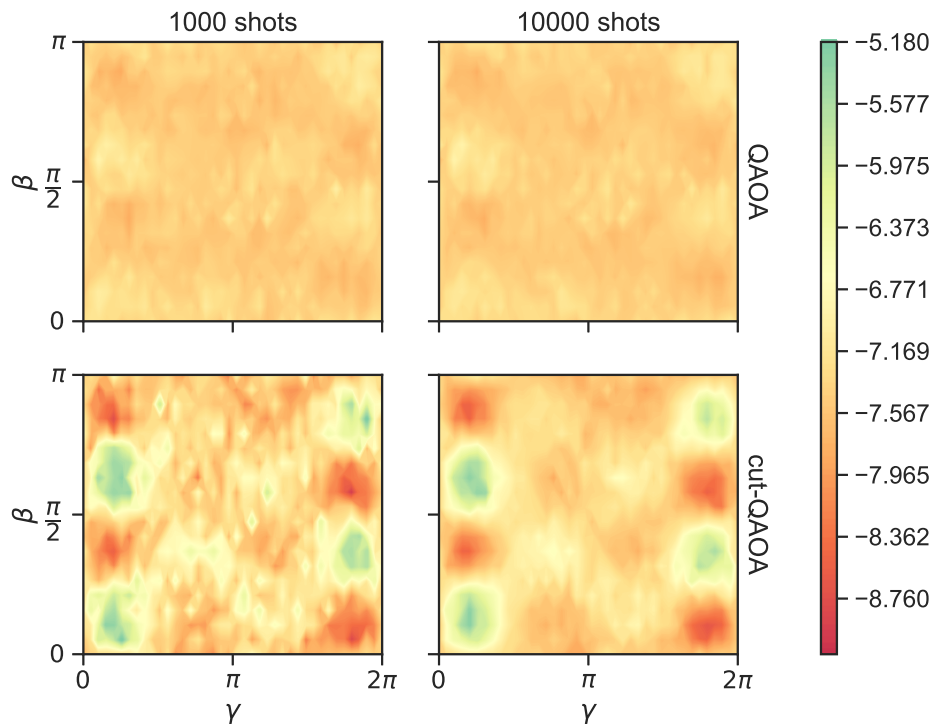}
     \caption{Computed objective functions on ibmq\_ehningen as parameter landscapes with the same color scale as the simulated objective function in \Cref{fig:exp1_sim_param_map}.} 
    \label{fig:exp1_param_maps}
\end{figure}

\Cref{fig:exp1_param_maps} visualizes the objective functions as parameter landscapes with 1000 and 10000 shots for QAOA and cut-QAOA, respectively, executed on the NISQ device \textit{ibmq\_ehningen}, a 27-qubit superconducting quantum device by IBM with their Falcon chip architecture.
All parameter landscapes are colorized with the same color scale as used for the simulated result in \Cref{fig:exp1_sim_param_map}.
It is evident that QAOA and cut-QAOA computed on ibmq\_ehningen achieve a smaller range of objective values than the simulated objective function~(see \Cref{fig:exp1_sim_param_map,fig:exp1_param_maps}).
However, the QAOA ansatz achieves a substantially smaller range of objective values than the cut-QAOA.
As seen in \Cref{fig:exp1_param_maps}, the maxima and minima of the cut-QAOA emerge more clearly from the plane compared to QAOA.
While with QAOA no noticeable change in the objective function can be seen between 1000 and 10000 shots, with cut-QAOA the objective function becomes noticeably smoother.

\begin{figure}
     \centering
     \begin{subfigure}[t]{0.49\linewidth}
         \centering
         \includegraphics[width=\linewidth]{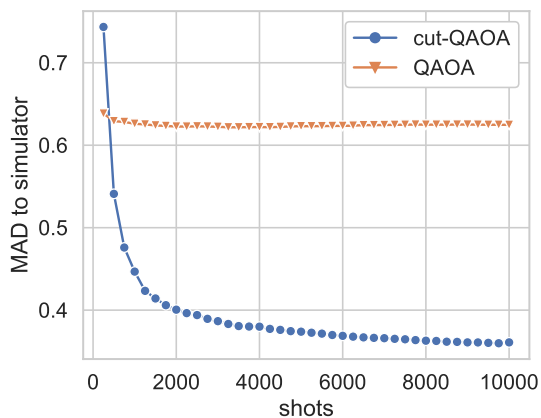}
         \caption{Mean absolute distance between simulator and quantum device}
         \label{fig:exp1_mean_absolute_distance}
     \end{subfigure}
     \hfill
     \begin{subfigure}[t]{0.49\linewidth}
         \centering
         \includegraphics[width=\linewidth]{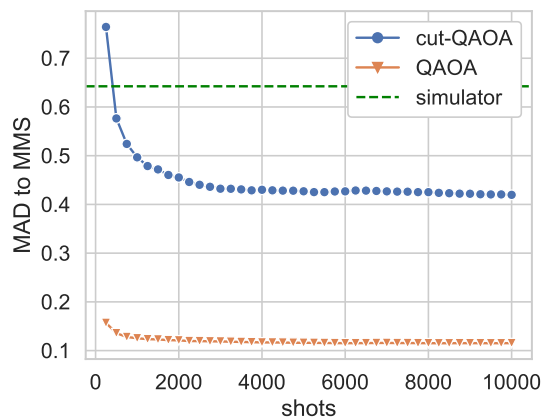}
         \caption{Mean absolute distance between quantum device and maximally mixed state}
         \label{fig:exp1_max_mixed_diff}
     \end{subfigure}
     \vspace{5mm}\\
     \begin{subfigure}[t]{0.49\linewidth}
         \centering
         \includegraphics[width=\linewidth]{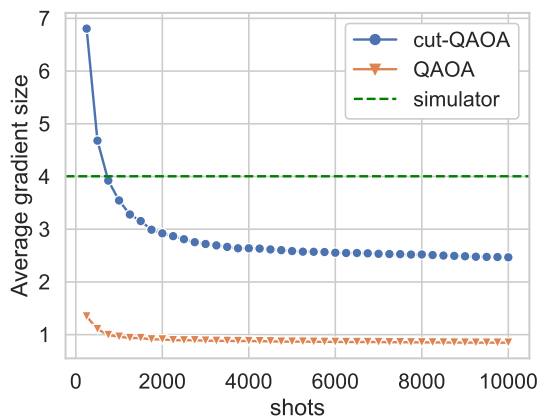}
         \caption{Average gradient size}
         \label{fig:exp1_gradient_length_avg}
     \end{subfigure}
     \hfill
     \begin{subfigure}[t]{0.49\linewidth}
         \centering
         \includegraphics[width=\linewidth]{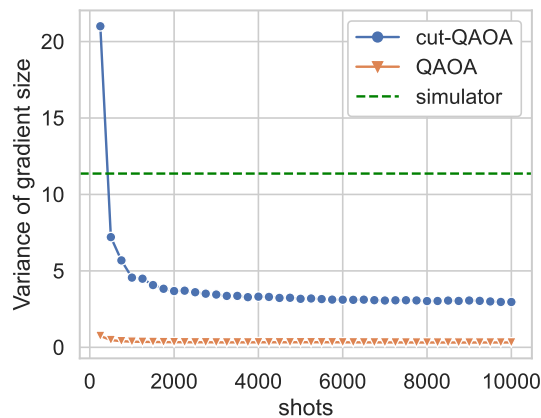}
         \caption{Variance of gradient size}
         \label{fig:exp1_gradient_length_var}
     \end{subfigure}
     \vspace{5mm}\\
     \hfill
     \begin{subfigure}[t]{0.49\linewidth}
         \centering
         \includegraphics[width=\linewidth]{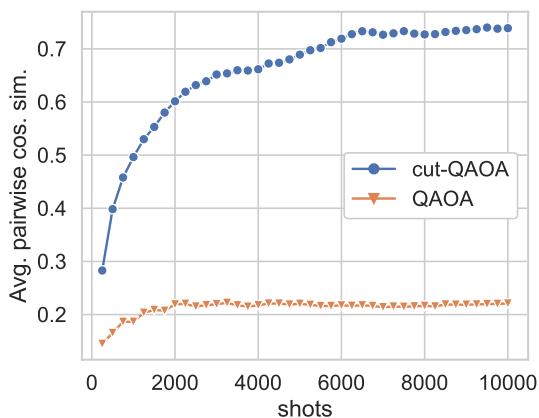}
         \caption{Average pairwise cosine similarity of gradients}
         \label{fig:exp1_similarity_avg}
     \end{subfigure}
     \hfill
     \vspace{2mm}\\
    \caption{Metrics for example graph as functions of shots}
    \label{fig:exp1_param_maps_metrics}
\end{figure}

In \Cref{fig:exp1_param_maps_metrics}, the introduced metrics for the objective functions above are plotted as a function of the number of shots.
It can be seen that the objective function of cut-QAOA is closer to the simulated objective function.
The MAD between cut-QAOA on ibmq\_ehningen and the simulation is significantly lower than for QAOA (\Cref{fig:exp1_mean_absolute_distance}).
Moreover, the deviation of cut-QAOA from the objective value of the MMS is substantially higher than the deviation of the QAOA ansatz~(\Cref{fig:exp1_max_mixed_diff}).
Furthermore, the more pronounced minima and maxima in the cut-QAOA objective function result in larger gradients with higher variance~(\Cref{fig:exp1_gradient_length_avg,fig:exp1_gradient_length_var}).
The average gradient size of the QAOA ansatz is significantly lower than with cutting, and the variance of gradient size is close to zero, indicating approximately equal size among the gradients.
In contrast, the objective function of cut-QAOA has strikingly increased gradients of higher variance.
This implies more salient structures in the computed objective function.
However, cut-QAOA still does not achieve the same magnitude and variance of gradients as the simulator.
Apart from the gradient size, the gradients of the cut-QAOA ansatz have a significantly higher cosine similarity value than the QAOA ansatz on the quantum devices (\Cref{fig:exp1_similarity_avg}).
Consequently, the mismatch in the gradient direction is smaller, and thus, the optimum and the optimization path to the optimum deviates less from the noise-free objective function.

All metrics as functions of the number of shots show the same pattern in \Cref{fig:exp1_param_maps_metrics}.
They are approaching a plateau. 
A reasonable explanation for this phenomenon is \textit{shot noise} which refers to statistical sampling errors that occur when estimating the result distribution of a quantum circuit with a finite number of shots~\cite{Knill2007}.
With an increasing number of shots, the sampled result distribution of a quantum circuit converges to its expected distribution.
This behavior manifests itself in decreasing values for all metrics as the number of shots increases except for gradient similarity.
There, the values increase as they approach a plateau.
The more the curves approach the plateaus, the less shot noise matters, but NISQ errors remain.
While this state is reached with QAOA with relatively few shots, cut-QAOA needs significantly more shots and thus is considerably more affected by shot noise.
This effect also fits with the fact that cut-QAOA must distribute the number of shots among all subcircuits, and thus, each individual subcircuit receives significantly fewer shots.

\subsubsection{Evaluation using a set of random graphs}

\begin{figure}
    \centering
    \includegraphics[width=\linewidth]{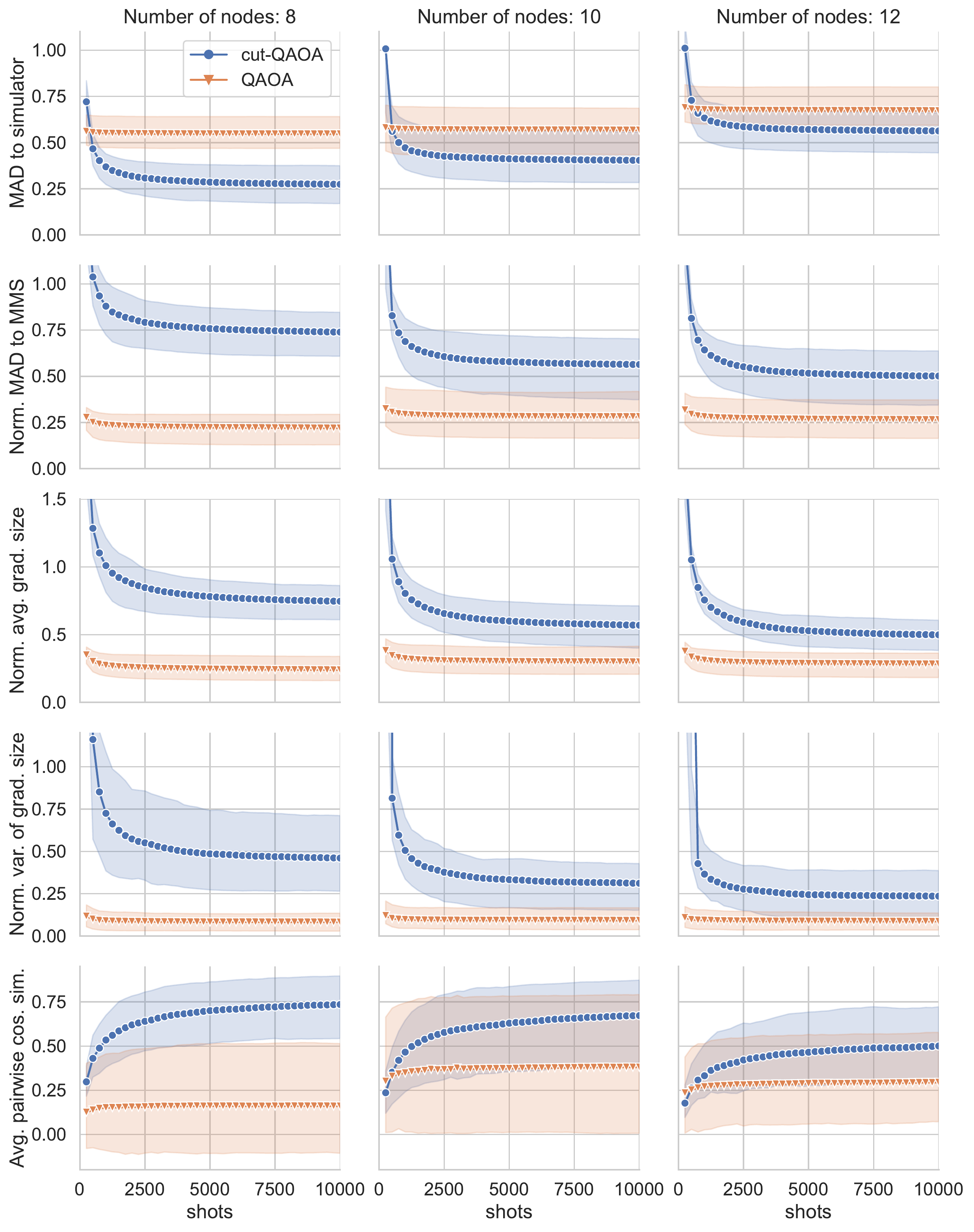}
    \caption{Mean value of a metric surrounded with a percentile interval of 80\%.}
    \label{fig:exp1_all_metrics}
\end{figure}

\begin{table}
    \begin{subtable}[h]{\textwidth}
        \centering
        \begin{tabular}{ l || l | l | l || l | l | l }
        &\multicolumn{3}{c||}{Exp. 1} & \multicolumn{3}{c}{Exp. 2}\\
        \hline
        Nodes & Count & Min Edges & Max Edges & Count & Min Edges & Max Edges \\
        \hline  
        8 & 18 & 8 & 14 &16 & 8 & 14\\
        10 & 25 & 10 & 16 & 15 & 11 & 16\\
        12 & 18 & 13 & 21 & 16 & 16 & 21
       \end{tabular}
       \caption{Statistics about the graphs in the dataset.}
       \label{tab:week1}
    \end{subtable}
    \newline
    \vspace*{0.4 cm}
    \newline
    \begin{subtable}[h]{\textwidth}
        \centering
        \begin{tabular}{l | l | l}
        Quantum device & Exp. 1 & Exp. 2 \\
        \hline 
        ibmq\_ehningen & 39 & 31\\
        ibmq\_mumbai & 10 & 6\\
        ibmq\_kolkata & 6 & 5\\
        ibmq\_guadalupe & 5 & 5\\
        ibmq\_montreal & 1 & 0
        \end{tabular}
        \caption{Used quantum devices for the experiemts.}
        \label{tab:week2}
     \end{subtable}
     \caption{Statistics about the data set: (a) graphs and (b) quantum devices.}
     \label{tab:dataset}
\end{table}

Our data set includes 61 randomly generated graphs as listed in \Cref{tab:dataset}.
The individual graphs can be viewed in the published dataset~\cite{codeanddata}.
\Cref{fig:exp1_all_metrics} shows each metric as a function of the number of shots for the random graphs from the data set.
For each metric and QAOA variant, the mean value is plotted as a line.
The shaded area around the mean represents a percentile interval of 80\% of the data. 
Thus, the highest and lowest 10\% of the values are not visualized.
To ensure better comparability of the metrics between problem instances, the MAD to MMS, average gradient size and variance of gradient size were normalized by dividing by the metric value of the simulator.
The other two metrics were not normalized further, since they already incorporated the simulator. 
The averaged values of the data set show the same patterns as observed for the example graph.
First, all metrics approach a plateau value with an increasing number of shots. 
As in the example above, cut-QAOA converges significantly slower on average and starts further away from its plateau value than QAOA for all metrics.
Both facts indicate that cut-QAOA is more susceptible to shot noise.

Moreover, the computed objective function of the cut-QAOA ansatz is closer to the simulated objective function and deviates more from the objective value of the maximally mixed state.
Furthermore, its gradients are of larger size and have a higher variance in their magnitude.
Additionally, the gradients of the objective function of the cut-QAOA are more similar to the simulated gradients.
Although \Cref{fig:exp1_all_metrics} shows only averaged metrics, the individual results of the experiments also follow these patterns and resemble the results of the example graph above (\Cref{fig:exp1_graph}). 
Moreover, while cut-QAOA performs better regarding the metrics, it tends to have a slightly higher spread in metric values across the problem instances except for gradient similarity, where the variance for QAOA is significantly larger.

However, these patterns are more or less pronounced depending on the number of nodes in the generated graph.
With an increasing number of nodes, and thus large quantum circuits, the advantage of cut-QAOA in the metrics decreases and approaches the value of QAOA.
For this, a probable reason is that with increasing graph size, the QAOA ansätze grow in depth and width, and so do the subcircuits of cut-QAOA.
Thus, each subcircuit is more susceptible to the noise of the NISQ devices, and thus the combined result of cut-QAOA gets noisier.

\begin{mdframed}
    [
        skipabove=15pt,
        innerbottommargin=.3\baselineskip,
        rightmargin=0em,
        leftmargin=0em
    ]
    \textbf{Major observations:}
    \begin{itemize}
        \item Compared to QAOA without cutting, the objective function of cut-QAOA computed on NISQ devices is closer to the simulated objective function.
        \item It deviates more from the objective value of the maximally mixed state.
        \item It has larger gradients with a higher variance that are more similar to the simulated gradients.
        \item However, cut-QAOA is more affected by shot noise.
    \end{itemize}
\end{mdframed}

\subsection{Results for experiment 2: QAOA execution}

Next, we evaluate whether these improvements in the computed objective function lead to better results in QAOA to clarify SRQ~2, i.e., lower expectation values and better MaxCut solutions.
We extracted the shot results of the ansatz for the optimized parameters $\beta^{*}, \gamma^{*}$ for each run of QAOA and cut-QAOA and considered the obtained expectation value $\Braket{H_{C}}_{\beta^{*}, \gamma^{*}}$ and the most frequently sampled state $z^{*}$.
To compare these results among different MaxCut problem instances, we normalize the result with the optimal objective value $\mathcal{C}_{\text{opt}}$. 
This yields the \textit{expectation ratio}

\begin{equation}
    r_{\text{exp}} = \dfrac{\Braket{H_{C}}_{\beta^{*}, \gamma^{*}}}{\mathcal{C}_{\text{opt}}}
\end{equation}

\noindent
and the \textit{approximation ratio}

\begin{equation}
    r_{\text{appr}} = \dfrac{\mathcal{C}(z^{*})}{\mathcal{C}_{\text{opt}}}.
\end{equation}

\noindent
While QAOA optimizes the expectation value, the state most frequently sampled is usually taken as the solution to the problem.

In addition to the computed solution with the QAOA variants on the quantum devices, we classically draw samples randomly from a uniform distribution of the solutions, i.e., we classically simulated sampling the MMS.
This sample serves as a benchmark to show whether the QAOA variant on the quantum device performs better than random sampling.
Furthermore, we executed QAOA on a noise-free simulator to obtain the expectation ratio and approximation ratio in an ideal, noise-free scenario. 
These noise-free solutions provide a reference point for evaluating the impact of noise and imperfections in the executions of the QAOA variants on the NISQ devices.

\begin{figure}
    \centering
    \includegraphics[width=\linewidth]{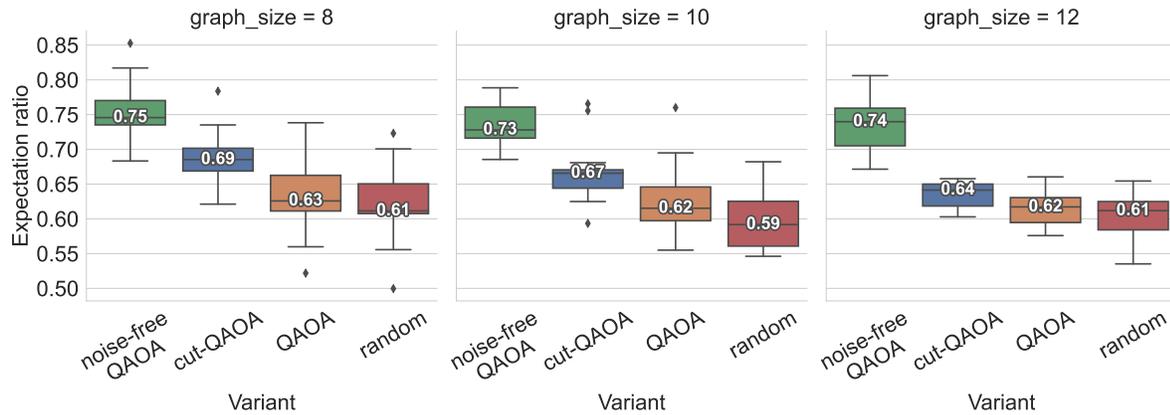}
    \caption{Boxplots of the expectation ratio for each variant with median value. The whiskers extend from the box by 1.5 times the interquartile range, and points outside this range are drawn as outliers~\cite{Dekking2005}.}
    \label{fig:exp2_expectation_boxplot}
\end{figure}

\begin{figure}
    \centering
    \includegraphics[width=\linewidth]{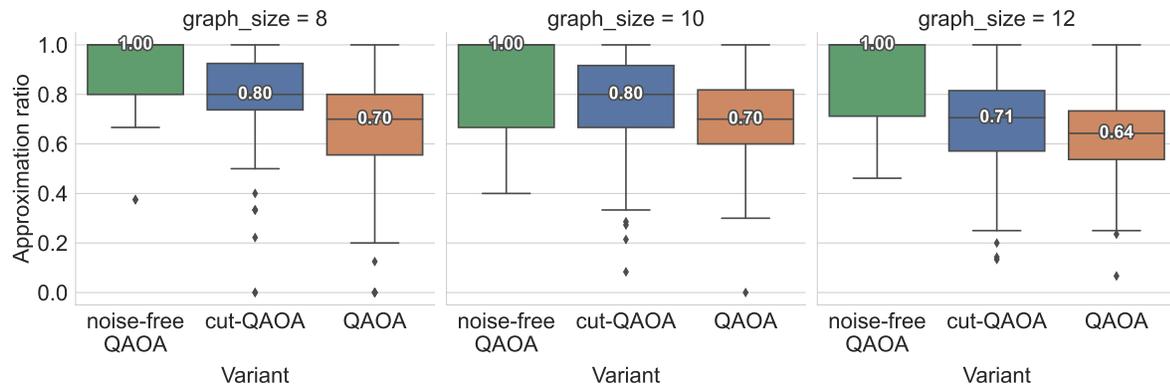}
    \caption{Boxplots of the approximation ratio for each QAOA variant with median value.}
    \label{fig:exp2_appr_ratio_boxplot}
\end{figure}

We conducted this experiment on a subset of graphs from experiment 1.
The data set includes 47 graphs as described in \cref{tab:dataset}.
The obtained expectation values can be seen in \Cref{fig:exp2_expectation_boxplot} for the different QAOA variants and graph sizes.
Cut-QAOA achieves the highest expectation ratio, while QAOA performs hardly better than random sampling.
However, as the graphs get larger, the advantage of cut-QAOA gets smaller, and it also approaches the expectation ratio of QAOA.
Considering the results from experiment 1, the computed objective function gets noisier, and thus, it is less accurate and flattens.
Therefore, it is harder for the classical optimizer to find optimal parameters, and even for optimal parameters, the sampled noisy solutions yield a lower expectation value in general.
The fact that QAOA without cutting is only marginally different from random sampling suggests that noise in its computation prevails, rendering the results impractical.
The better expectation values of cut-QAOA also translate to solutions of higher approximation ratio, as can be seen in \Cref{fig:exp2_appr_ratio_boxplot}.
Cut-QAOA follows the previously established pattern and approaches the approximation ratio of QAOA with increasing graph size.

\begin{mdframed}
    [
        skipabove=15pt,
        innerbottommargin=.3\baselineskip,
        rightmargin=0em,
        leftmargin=0em
    ]
    \textbf{Major observations:}
    \begin{itemize}
        \item QAOA without cutting performs only marginally better than random sampling.
        \item Cut-QAOA achieves noticeably better expectation values than QAOA, but this advantage decreases with increasing size of the graphs.
        \item Higher expectation values of cut-QAOA translate to solutions with noticeably higher approximation ratios.
    \end{itemize}
\end{mdframed}

\section{Discussion}\label{sec:discussion}

The observed result that quantum circuit cutting applied in QAOA for the MaxCut problem produces execution results less affected by noise are in line with evaluations of circuit cutting on benchmark circuits~\cite{Tang2021,Ayral2021,Ying2022}.
The reasoning is that there is a range in the size of the subcircuits where their execution is significantly less noisy than the execution of the original circuit.
Thus, the classical, noise-free recombination procedure can estimate a less noisy result for the original circuit based on the less noisy subcircuit results.
While our work focuses only on the decomposition of the circuit into two fragments that bisects the number of qubits in the subcircuits, another work shows that by decomposing the circuit into more fragments, the effect is further enhanced~\cite{Ayral2021}. 
However, each additional cut of the circuit increases the number of subcircuits and the shots required to sample them in the worst case exponentially, and thus choosing the number of cuts and fragments is a tradeoff between smaller circuits and additional cost.
This additional cost is also manifested in our experiments by the higher shot noise, i.e., more shots are needed for convergence.
This observation is consistent with theoretical considerations of circuit cutting in other work~\cite{Piveteau2022,Mitarai2021a} and, in general, the applicability of circuit cutting to suppress errors and tackle NIBPs depends on structural properties of the circuits that determine the number of cuts and therefore the required overhead.
That means, to effectively scale cut-QAOA to larger problem instances while keeping computational overhead manageable, it is crucial to ensure that the number of required cuts to partition the circuit grows slowly compared to the problem size. 
By doing so, we can delay the exponential increase in overhead, thereby expanding the number of problem instances solvable in an acceptable time.
The corresponding circuits of these problem instances should exhibit a specific cluster structure that is characterized by high two-qubit gate connectivity among qubits within the same cluster and low connectivity between clusters. 
Leveraging this structure allows for efficient circuit partitioning with a small number of cuts relative to the problem size.

\subsection{Limitations}
While our findings offer valuable perspectives and potential avenues for further investigation, their generalizability may be limited due to the influence of various factors, such as the selection of the problem, the problem instances, and the experimental setup. 
Even though we focus on solving the MaxCut problem with QAOA in our work, the application of circuit cutting is independent of it since it operates at the circuit level and is detached from the underlying problem and algorithm.
Moreover, the experiments were only conducted with a limited set of graphs due to the restricted availability of NISQ devices and execution time on these.
All investigated graphs have either eight, ten, or twelve nodes since, for graphs with more nodes, the noise dominates even with cut-QAOA.
To control the number of cuts per problem instance in the experiments, each graph in the dataset has a particular structure: it consists of two equal-sized subgraphs connected by two edges.
The choice of two edges has two reasons.
On the one hand, the solution to the MaxCut problem cannot be generated in general by combining the solutions of the two subgraphs.
On the other hand, the used circuit cutting approach must execute only 32 subcircuits for the two cuts.
Other work focuses on experiments with more than two cuts~\cite{Ying2022,Ayral2021} and also on significantly reducing the number of required subcircuits~\cite{Marshall2022}.
Regarding the selection of the NISQ devices, the experiments were conducted on multiple devices of IBM's Falcon chip architecture.
Furthermore, the calibration of the devices changes periodically and cannot be controlled by the user.
This harms the reproducibility of the experiments.
To compensate for this, we have included all calibration data of the quantum devices in the dataset to improve the traceability of results~\cite{codeanddata}.
Despite all these limitations, we obtained comparable results for the different graphs, NISQ devices, and their calibrations that show the same patterns concerning the application of circuit cutting.

\subsection{Correlation between objective function metrics and QAOA result}

\begin{figure}
    \centering
    \includegraphics[width=\linewidth]{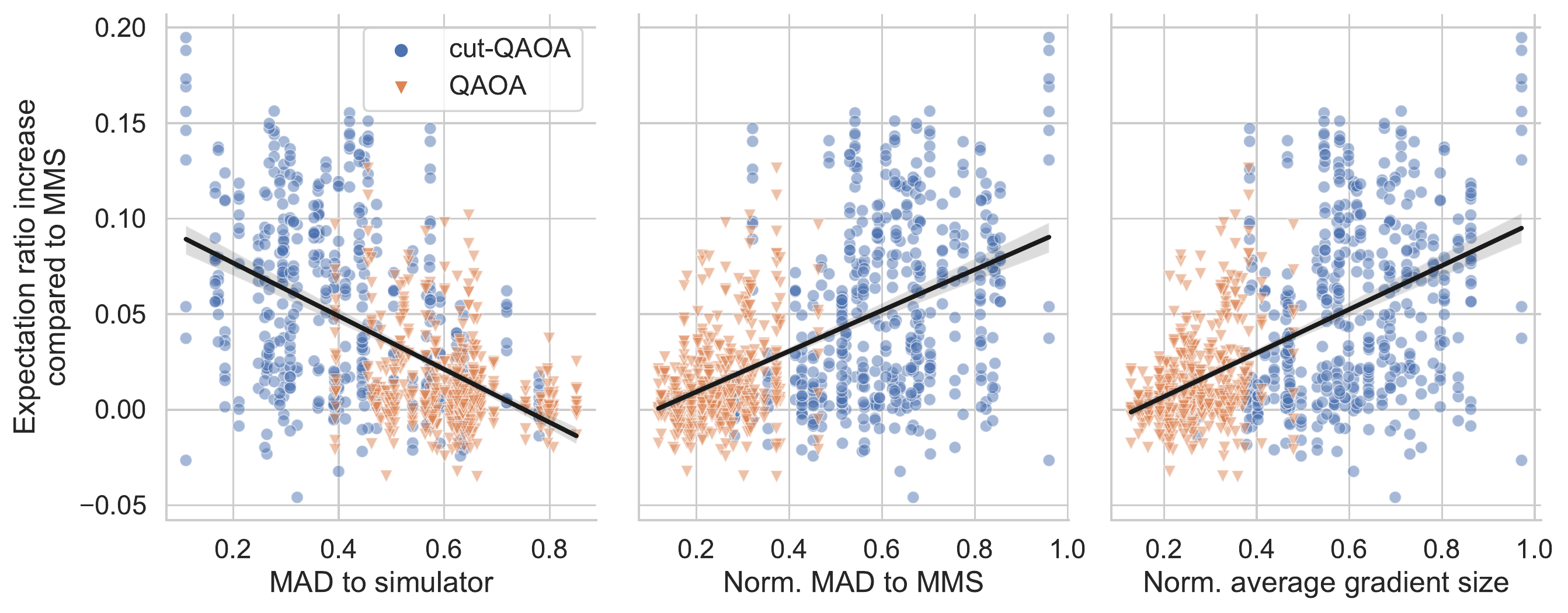}
    \caption{Correlation between metric and the increase in expectation ratio compared to the MMS. The black line visualizes the regression line surrounded by a confidence interval of 0.95 for the regression estimate.}
    \label{fig:correlation}
\end{figure}

The connection between the computed objective function's quality and the quality of the result of the algorithm becomes evident when combining the data from the two conducted experiments.
We can see a correlation between the introduced metrics of objective functions and the expectation and approximation ratio of the corresponding QAOA results. 
QAOA tends to produce better results for objective functions that perform better concerning the metrics.
\Cref{fig:correlation} visualizes the correlations between three metrics and the increase in expectation ratio compared to the MMS, i.e., $\left(\Braket{H_{C}}_{\beta^{*}, \gamma^{*}} - \frac{1}{2^{n}}\tr\left[H_{C}\right]\right) \mathcal{C}_{\text{opt}}^{-1}$. 
The correlation of other metrics can be found in the data set~\cite{codeanddata}.
While there are always runs where no improvement was achieved, e.g., due to poor initial parameters and local optima, the advantage in expected value increases as the metrics improve.
This positive correlation supports the relevance of the metrics used, and in addition, it fortifies the assumption that an improved objective function aids the classical optimizer in finding better solutions.
Furthermore, these results were obtained, although the COBYLA optimizer in the default configuration was employed, i.e., without adjusting its hyperparameters to the optimization task.
Further tuning of the optimizer to the underlying problem and noise can lead to better results~\cite{Lavrijsen2020}.

\section{Summary and Future Work}\label{sec:conclusion}

In this paper, we applied circuit cutting in QAOA to solve the MaxCut problem on current NISQ devices.
To answer SRQ~1, which deals with the influence of circuit cutting on the objective function, our results provide insight that circuit cutting suppresses the effects of noise in the computed objective function on NISQ devices.
We observe objective functions computed with circuit cutting that are less affected by NIBPs.
They are closer to the noise-free result, are less concentrated around the objective value of the maximally mixed state, and have larger gradients with more variance that are more similar to the expected gradients than the computed objective functions without circuit cutting.
However, their computation with circuit cutting exhibits more shot noise, i.e., more shots are needed for convergence. 
Regarding SRQ~2, it can be observed in our experiments that QAOA with circuit cutting obtains better solutions on NISQ devices by classical optimization of the cut ansatz than without cutting.
It achieves better expectation values and produces better solutions, i.e., its most frequently sampled state, with a higher approximation ratio.

Part of our future work will be to investigate how large the range of problem instances is where advantages can be obtained with circuit cutting on NISQ devices. 
Small problem instances at the lower end of the range can also be computed with low noise without cutting, and with large problem instances at the upper end, already too many errors occur in the subcircuit's computation.

Moreover, as circuit cutting operates at the level of quantum circuits, it is to be expected that the results generalize to other problems and VQAs.
Furthermore, other decomposition schemes in VQAs~\cite{Li2021,Zhou2022,Shaydulin2019,Shaydulin2019a,Tomesh2021} that reduce the size of executed quantum circuits may exhibit similar effects.
Both will be evaluated as part of our future work.

\appendix
\section{}\label{ap:rzz_cut}
The operation of $R_{ZZ}(\gamma)$ can  be expressed as follows:
\begin{equation}
    R_{ZZ}(\gamma) = \exp\left(\frac{-i \gamma Z \otimes Z}{2}\right) = \cos\left(\frac{\gamma}{2}\right) I \otimes I - i \sin\left(\frac{\gamma}{2}\right) Z \otimes Z
\end{equation}
Apply $R_{ZZ}(\gamma)$ to an arbitrary state $\ket{\varphi_{1}\varphi_{0}}$ with density operator $ \rho = \ket{\varphi_{1}\varphi_{0}}\bra{\varphi_{1}\varphi_{0}}$:
\begin{equation}
\begin{split}
    R_{ZZ}(\gamma)\rho R_{ZZ}(\gamma)^{\dagger} &=  \cos^{2}\left(\frac{\gamma}{2}\right) I^{\otimes2}\rho I^{\otimes2} + \sin^{2}\left(\frac{\gamma}{2}\right) Z^{\otimes2}\rho Z^{\otimes2} 
    \\&\quad+ i\cos\left(\frac{\gamma}{2}\right)\sin\left(\frac{\gamma}{2}\right)\left(I^{\otimes2}\rho Z^{\otimes2} - Z^{\otimes2}\rho I^{\otimes2}\right).
\end{split}
\end{equation}

\noindent
The operation $\left(I^{\otimes2}\rho Z^{\otimes2} - Z^{\otimes2}\rho I^{\otimes2}\right)$ can be implemented on a NISQ device as the following.
Let $ \rho_{0} = \ket{\varphi_{0}}\bra{\varphi_{0}}$ and $ \rho_{1} = \ket{\varphi_{1}}\bra{\varphi_{1}}$.
Thus, $\rho = \rho_{1} \otimes \rho_{0}$.
Let $P_{0} = \frac{I + Z}{2}$ and $P_{1} = \frac{I - Z}{2}$ be the projection on $\ket{0}$ and $\ket{1}$, respectively.
\begin{equation}
    A_{i} := R_Z\left(-\frac{\pi}{2}\right)\rho_{i} R_Z\left(-\frac{\pi}{2}\right)^{\dagger} - R_Z\left(\frac{\pi}{2}\right)\rho_{i} R_Z\left(\frac{\pi}{2}\right)^{\dagger}
\end{equation}
\begin{equation}
    B_{i} := P_{1}\rho_{i} P_{1}^{\dagger} - P_{0}\rho_{i} P_{0}^{\dagger}
\end{equation}
Putting it together:
\begin{equation}
    A_{0}\otimes B_{1} + B_{0}\otimes A_{1} = i\left(I^{\otimes2}\rho Z^{\otimes2} - Z^{\otimes2}\rho I^{\otimes2}\right)
\end{equation}
This results in the following:
\begin{equation}
    \begin{split}
    R_{ZZ}(\gamma)\rho R_{ZZ}(\gamma)^{\dagger} &=  \cos^{2}\left(\frac{\gamma}{2}\right) I^{\otimes2}\rho I^{\otimes2} + \sin^{2}\left(\frac{\gamma}{2}\right) Z^{\otimes2}\rho Z^{\otimes2} 
    \\&\quad+ \cos\left(\frac{\gamma}{2}\right)\sin\left(\frac{\gamma}{2}\right)\left(A_{0}\otimes B_{1} + B_{0}\otimes A_{1}\right)
\end{split}
\end{equation}

\section{}\label{ap:rz_decomposition}
The operation of $R_{Z}(\gamma)$ can  be expressed as follows:
\begin{equation}
    R_{Z}(\gamma) = \exp\left(\frac{-i \gamma Z}{2}\right) = \cos\left(\frac{\gamma}{2}\right) I  - i \sin\left(\frac{\gamma}{2}\right) Z. 
\end{equation}
Thus, the corresponding superoperator is 
\begin{equation}
    \superop(R_{Z}(\gamma))\rho  =  \cos^{2}\left(\frac{\gamma}{2}\right) \superop(I)\rho  + \sin^{2}\left(\frac{\gamma}{2}\right) \superop(Z)\rho 
    + i\cos\left(\frac{\gamma}{2}\right)\sin\left(\frac{\gamma}{2}\right)\left(I\rho Z - Z\rho I\right).
\end{equation}
For angle $-\gamma$ it holds
\begin{equation}
    \superop(R_{Z}(-\gamma))\rho =  \cos^{2}\left(\frac{\gamma}{2}\right) \superop(I)\rho + \sin^{2}\left(\frac{\gamma}{2}\right) \superop(Z)\rho 
    - i\cos\left(\frac{\gamma}{2}\right)\sin\left(\frac{\gamma}{2}\right)\left(I\rho Z - Z\rho I\right).
\end{equation}
Together:
\begin{equation}
    \superop(R_{Z}(-\gamma))\rho  - \superop(R_{Z}(\gamma))\rho 
     = - 2 i\cos\left(\frac{\gamma}{2}\right)\sin\left(\frac{\gamma}{2}\right)\left(I\rho Z - Z\rho I\right)\\
\end{equation}
It follows:
\begin{equation}
    \superop(R_{Z}(-\gamma))  - \superop(R_{Z}(\gamma))
     = 2\cos^{2}\left(\frac{\gamma}{2}\right) \superop(I) + 2\sin^{2}\left(\frac{\gamma}{2}\right) \superop(Z) - 2\superop(R_{Z}(\gamma))
\end{equation}

\ack 
This work was partially funded by the BMWK projects \textit{PlanQK} (01MK20005N), \textit{EniQmA} (01MQ22007B), and \textit{SeQuenC} (01MQ22009B).

\section*{Data availability}
The data that support the findings of this study are openly available at the following URL: \url{https://doi.org/10.18419/darus-3124}

\bibliographystyle{unsrt}

\section*{References}

\bibliography{bibliography.bib}

\vspace{1cm}
\noindent
All links were last followed on July 10, 2023.

\end{document}